%% file: new_main.tex
\documentclass[acmsmall, authorversion, nonacm]{acmart}
\AtBeginDocument{%
  \providecommand\BibTeX{{%
    \normalfont B\kern-0.5em{\scshape i\kern-0.25em b}\kern-0.8em\TeX}}}

\setcopyright{none}
\copyrightyear{2022}
\acmYear{2022}
\acmDOI{none}

\acmJournal{JACM}
\acmVolume{TBD}
\acmNumber{TBdD}
\acmArticle{}
\acmMonth{1}

\usepackage{graphicx}
\usepackage{slashbox}
\usepackage{caption}
\usepackage{subcaption}
\usepackage{xcolor}
\usepackage{cite}
\usepackage{hyperref}

\newtheorem{definition}{Definition} % definition numbers are dependent on theorem numbers
\newtheorem{theorem}{Theorem} % same for example numbers
\newtheorem{remark}{Remark}
\newtheorem{lemma}{Lemma}
\newtheorem{corollary}{Corollary}
\newtheorem{example}{Example}
\newtheorem{proposition}{Proposition}
\newenvironment{myexpcont}
{\addtocounter{example}{-1}\begin{example}{\textit{\textbf{{(continued)}}}}}
  {\end{example}}
\newenvironment{myexpcont2}
{\addtocounter{example}{-2}\begin{example}{\textit{\textbf{{(continued)}}}}}
  {\end{example}}

\begin{document}

\title{Risk of Stochastic Systems for Temporal Logic Specifications}

\author{Lars Lindemann}
\email{larsl@seas.upenn.edu}
\orcid{0000-0003-3430-6625}
\author{Lejun Jiang}
\orcid{0000-0003-3430-6625}
\email{lejunj@seas.upenn.edu}
\author{Nikolai Matni}
\orcid{0000-0003-3430-6625}
\email{nmatni@seas.upenn.edu}
\author{George J. Pappas}
\orcid{0000-0003-3430-6625}
\email{pappasg@seas.upenn.edu}
\affiliation{%
  \institution{University of Pennsylvania}
  \streetaddress{200 South 33rd Street}
  \city{Philadelphia}
  \state{Pennsylvania}
  \country{USA}
  \postcode{19104}
}

\renewcommand{\shortauthors}{Lindemann, et al.}

\begin{abstract}
The wide availability of data coupled with the computational advances in artificial intelligence and machine learning promise to enable many future technologies such as autonomous driving. While there has been a variety of successful demonstrations of these technologies, critical system failures have repeatedly been reported. Even if rare, such system failures pose a serious barrier to adoption without a rigorous risk assessment. This paper presents a framework for the \emph{systematic and rigorous risk verification} of systems. We consider a wide range of system specifications formulated in signal temporal logic (STL) and model the system as a stochastic process, permitting discrete-time and continuous-time stochastic processes. We then define the STL robustness risk as \emph{the risk of lacking robustness against failure}. This definition is motivated as system failures are often caused by missing robustness to modeling errors, system disturbances, and  distribution shifts in the underlying data generating process. Within the definition, we permit general classes of risk measures and  focus on tail risk measures such as the value-at-risk and the conditional value-at-risk. While the STL robustness risk is in general hard to compute, we propose the approximate STL robustness risk as a more tractable notion that upper bounds the STL robustness risk. We show how the approximate STL robustness risk can accurately be estimated from system trajectory data. For  discrete-time stochastic processes, we show under which conditions the approximate STL robustness risk can even be  computed exactly. We illustrate our verification algorithm in the autonomous driving simulator CARLA and show how a least risky controller  can be selected among four  neural network lane keeping controllers for five meaningful system specifications. 
\end{abstract}


%\includegraphics[scale=0.05]{figures/carla_}
%\caption{figure caption}
%\Description{figure description}

\begin{teaserfigure}
\centering
\includegraphics[scale=0.061]{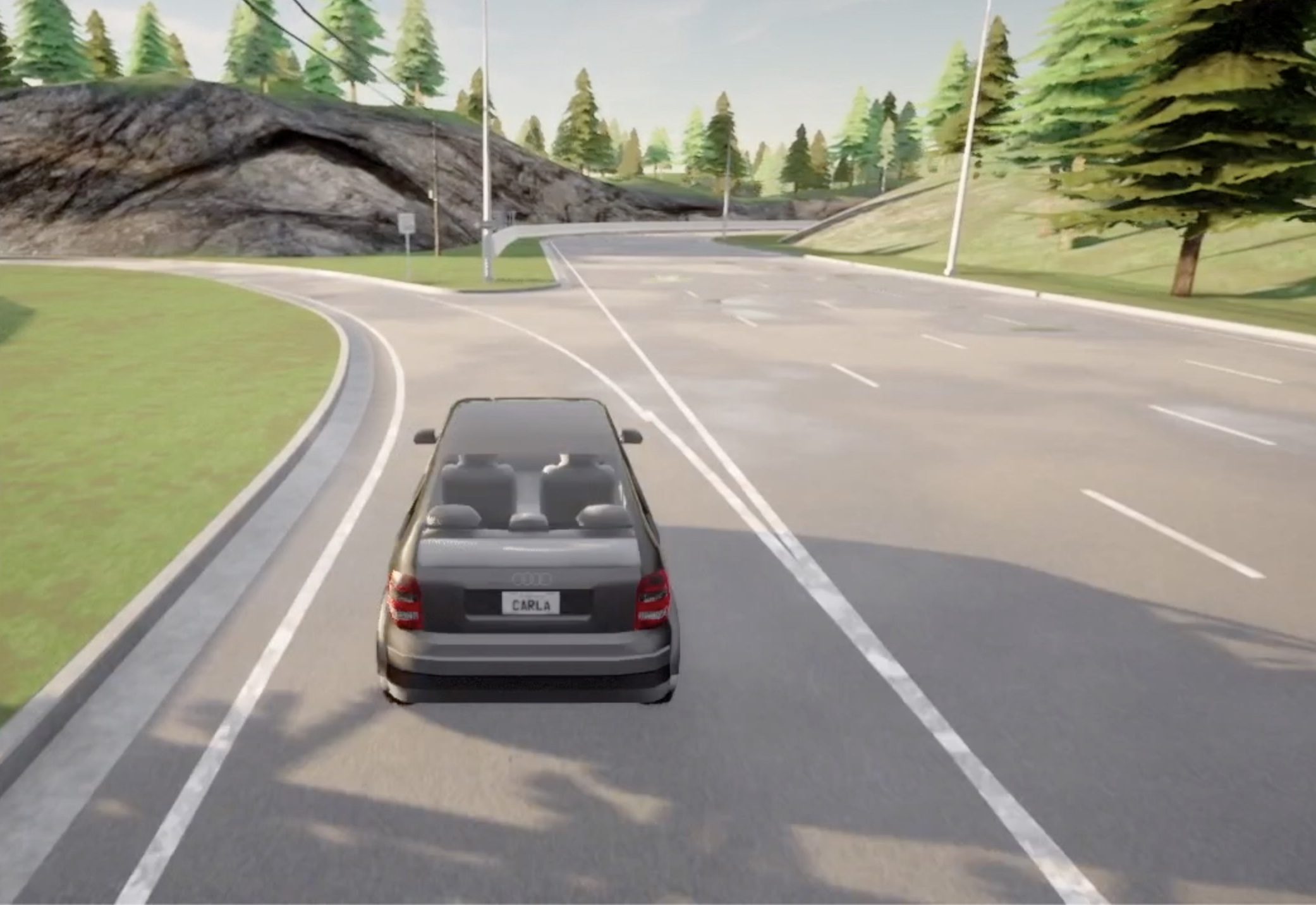}\hspace{0.1cm}
\includegraphics[scale=0.0625]{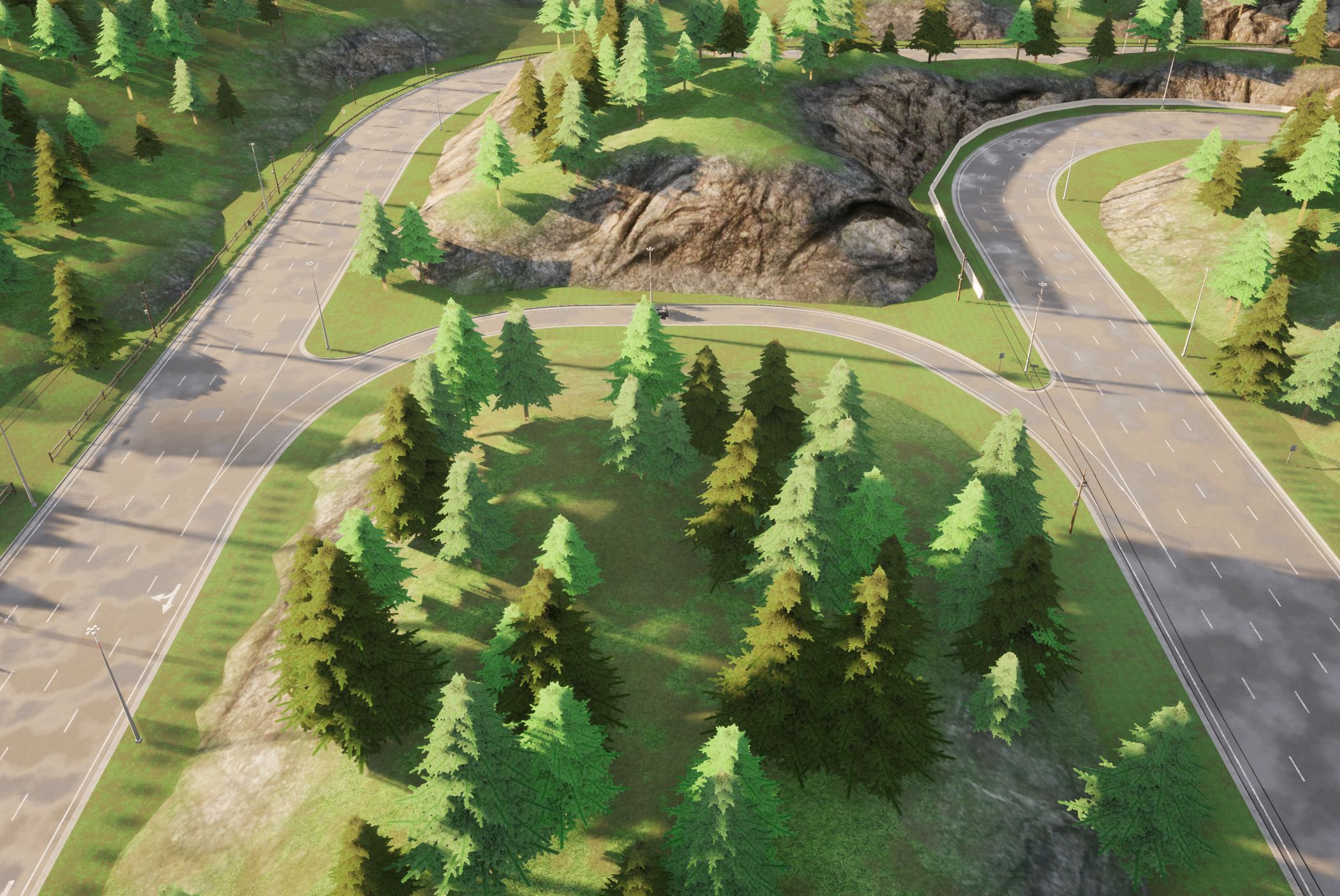}\hspace{0.1cm}
\includegraphics[scale=0.365]{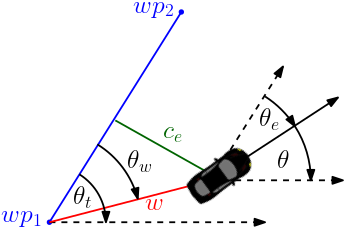}
\caption{Left: Simulation environment in the autonomous driving simulator CARLA. Middle: Double left turn on which we evaluate four trained neural network lane keeping controllers. Right: Cross-track error $c_e$ and orientation error $\theta_e$ used for risk verification of the neural network controllers. }
\label{fig:CARLA_}
\end{teaserfigure}

\maketitle

\input{chapter/introduction}
\input{chapter/background}

\input{chapter/problem_formulation}

\input{chapter/main}

\input{chapter/simulations}

\input{chapter/conclusion}

\begin{acks}
This research was supported by NSF award CPS-2038873 and NSF CAREER award ECCS-2045834, and a Google Research Scholar award.

\end{acks}

\bibliographystyle{ACM-Reference-Format}
\bibliography{literature}

\appendix
\input{chapter/appendix}

\end{document}

%% file: chapter/introduction.tex
\section{Introduction}
\label{sec:introduction}

Over the next decade, large amounts of data
will be generated and stored as devices that perceive and control the  world become more affordable and available. Impressive demonstrations of data-driven and machine learning enabled technologies exist already today, e.g., robotic manipulation \citep{levine2016end}, solving games \citep{silver2018general,mnih2015human}, and autonomous driving \citep{dosovitskiy2017carla}. However, occasionally occurring system failures impede the use of these technologies particularly when system safety is a concern. For instance, neural networks, frequently used for perception and control in autonomous systems, are known to be fragile and non-robust \citep{goodfellow2014explaining,su2019one}. Especially the problem of long tails in training data distributions poses challenges, e.g., natural variations in weather and lightning conditions \citep{robey2020model}.  

Moving forward, we expect that system failures appear less frequently due to advancing technologies -- nonetheless,  algorithms for the systematic and rigorous \emph{risk verification} of such systems is needed. For instance, the National Transportation Safety Board emphasized in a statement in connection with an Uber accident from 2018  ``the need for safety risk management requirements for testing automated vehicles on public roads''  \citep{national2019inadequate}. In this paper, we show how to reason about the risk of systems that are modeled as stochastic processes. We consider a wide range of system specifications formulated in signal temporal logic (STL) \citep{maler2004monitoring,bartocci2018specification} and present a systematic way to quantify and compute the risk of a system lacking robustness against  failure.

\subsection{Related Work}

Depending on the research disciplines and applications, risk can have various interpretations. While risk is often defined as a failure probability, it can also be understood in more general terms as a metric defined over a cost distribution, e.g., the expected value or the variance of a distribution. We focus  on \emph{tail risk measures} to capture the rare yet costly events of a distribution. In particular, we consider the value-at-risk (VaR), i.e., quantiles of a distribution, and the conditional value-at-risk (CVaR) \citep{rockafellar2000optimization,rockafellar2002conditional}, i.e., the expected value over a quantile. Tail risk measures are more frequently being used in robotics and control applications where system safety is important  \citep{majumdar2020should}. 

{\textbf{Risk in control.}} Control design under risk objectives and constraints is increasingly been studied among control theorists as machine learning components integrated into closed-loop systems cause stochastic system uncertainty.  Oftentimes, the CVaR risk measure is used  to capture risk due its convexity and the property of being an upper bound to the VaR. For instance, the authors in \citep{samuelson2018safety} consider a stochastic optimal control problem with CVaR constraints over the distance to obstacles. Linear quadratic control under risk constraints was considered in \citep{tsiamis2021linear} to trade off risk and mean performance. A similar idea is followed for the risk constrained minimum mean squared error estimator in \citep{kalogerias2020better}. Risk-aware model predictive control was considered in \citep{singh2018framework,hyeon2020fast}, while \citep{schuurmans2020learning,coulson2020distributionally} present data-driven and distributionally robust model predictive controllers. Risk-aware control barrier functions for safe control synthesis were proposed in \citep{ahmadi2022risk}, while \citep{nyberg2021risk} demonstrates the use of risk in sampling-based planning. We remark that we view these works to be orthogonal to our paper as we provide a data-driven framework for the risk assessment under complex temporal logic specifications, and we hope to inform future control design strategies.

{\textbf{Stochastic system verification.}} System verification has a long history in complementing and informing the control design process of systems, e.g., using model checking \citep{baier,cassandras2009introduction}. When dealing with stochastic systems, system verification becomes computationally more challenging \citep{kwiatkowska2007stochastic}. Statistical model checking has recently gained attention by relying on availability of data instead of computation \citep{legay2019statistical,agha2018survey,zuliani2010bayesian,fan2017d}. Another line of work considers stochastic barrier functions for safety verification of dynamical systems \citep{prajna2007framework,jagtap2018temporal}.  The authors in \citep{jasour2021fast,jasour2021real} deal with the verification of stochastic dynamial systems during runtime. Motivated by the fragility and sensitivity of neural networks \citep{goodfellow2014explaining,su2019one}, a special focus has recently been on verifying neural networks in open-loop \citep{katz2019marabou,singh2019abstract} and closed-loop \citep{ivanov2019verisig}. We remark that our algorithms presented in this paper permit verification of general classes of systems, including systems with neural networks, as long as we can obtain data, e.g., from a simulator.  The guarantees obtained in  these previous works are either worst case guarantees or in terms of failure probabilities. Towards incorporating tail risk measures, the authors in \citep{chapman2019risk,chapman2019riska} propose a risk-aware safety analysis framework using the CVaR. We are instead interested in system verification under more complex temporal logic specifications and risk. 

{\textbf{Temporal logics.}} We use signal temporal logic to express a wide range of system specifications, e.g., surveillance (``visit regions A, B, and C every $10-60$ sec"), safety (``always between $5-25$ sec stay at least $1$ m away from region D"), and many others. For deterministic signals, STL allows to calculate the robustness by which a signal satisfies an STL specification. Particularly, the authors in \citep{fainekos2009robustness} proposed the robustness degree as the maximal tube around a  signal in which all signals satisfy the specification. The size of the tube consequently measures the robustness of this signal with respect to the specification. As the robustness degree is in general hard to calculate, the authors in \citep{fainekos2009robustness} proposed approximate yet easier to calculate robust semantics. Many forms of robust semantics have appeared  such as space and time robustness \citep{donze2}, the arithmetic-geometric mean robustness \citep{mehdipour}, the smooth cumulative robustness \citep{haghighi2019control}, averaged STL \citep{akazaki2015time}, and \citep{rodionova2016temporal} in which a connection with linear time-invariant filtering is established allowing to define various types of robust semantics. 

For stochastic signals, the authors in \citep{tiger2020incremental,li2017stochastic,kyriakis2019specification,SadighRSS16,jha2018safe} propose notions of probabilistic signal temporal logic in which chance constraints over predicates are considered, while the Boolean and temporal operators of STL are not changed. Similarly, notions of risk signal temporal logic have recently appeared in  \citep{lindemann2021reactive,safaoui2020control,li2022learning} by defining risk constraints over predicates while not changing the definitions of Boolean and temporal operators. In this paper, we instead define risk over the whole STL specification. The work in \citep{farahani2018shrinking} considers the probability of an STL specification being satisfied instead of using chance or risk constraints over predicates. The authors in \citep{wang2019statistical} consider hyperproperties in STL, i.e., properties between multiple system executions. More with a control synthesis focus and for the less expressive formalism of linear temporal logic, the authors  in \citep{bharadwaj2018synthesis,vasile2016control,lahijanian2015formal} consider control over belief spaces, while the authors in \citep{guo2018probabilistic} consider probabilistic satisfaction over Markov decision processes. Complementary to these works, \citep{baharisangari2021uncertainty,puranic2021learning} propose techniques to infer STL specifications from data towards explaining the underlying data.

{\textbf{Risk verification with temporal logics.}} In this paper, we quantify and compute the risk of lacking robustness against failure. We argue that the consideration of robustness in system verification is crucial and are particularly motivated by the fact that system failures are often caused by missing robustness to modeling errors, system disturbances, and distribution shifts in the underlying data generating process. The authors in \citep{anevlavis2022being} further highlight the importance of robustness in system verification. Probably closest to our paper are the works in \citep{salamati2020data,salamati2021data,jackson2021formal} and \citep{bartocci2013robustness,bartocci2015system}. In \citep{salamati2020data,salamati2021data,jackson2021formal}, the authors combine data-driven and model-based verification techniques to obtain information about the satisfaction probability of a partially known system. The authors in \citep{bartocci2013robustness,bartocci2015system} present a purely data-driven verification technique to estimate probabilities over robustness distributions of the system. Conceptually our work differs in two directions. First, we consider general risk measures to be able to focus on the tails of the robustness distribution. We also show how to estimate the robustness risk from data with high confidence. Second, we use the robustness degree as defined in \citep{fainekos2009robustness} to obtain  robustness distributions. This in fact allows us to obtain a precise geometric interpretation of risk. This paper is based on our previous work \citep{lindemann2021stl}. We here permit continuous-time stochastic processes and the CVaR as a risk measure. We also show under which conditions the STL robustness risk can exactly be calculated, while presenting exhaustive simulations within the autonomous driving simulator CARLA \citep{dosovitskiy2017carla}.

\subsection{Contributions and Paper Outline}

Our general goal is to analyze the robustness of stochastic processes, and to quantify and compute the \emph{risk of a system lacking robustness against system failure}.  We make the following contributions:

\begin{itemize}
    \item We consider discrete-time and continuous-time stochastic processes and show under which conditions the robust semantics and the robustness degree of STL are random variables. This enables us to define risk over these quantities.
    \item We define the STL robustness risk as the risk of a system lacking robustness against failure of an STL specification. The definition permits general classes of risk measures and has a precise geometric interpretation in terms of the size of permissible disturbances. We also define the approximate STL robustness risk as a computationally tractable upper bound of the STL robustness risk.
    \item  For the VaR and the CVaR, we show how the approximate STL robustness risk can be  estimated from system trajectory data. Importantly, no particular restriction on the distribution of the stochastic process has to be made. For discrete-time stochastic processes with a discrete state space, we show how the approximate STL robustness risk can even be computed exactly. 
    \item We estimate the risk of four neural network lane keeping controllers within the  autonomous driving simulator CARLA. We show how to find the least risky controller.
\end{itemize}

In Section \ref{sec:backgound}, we present background on signal temporal logic, stochastic processes, and risk measures. In Section \ref{risskk}, we define the STL robustness risk and the STL approximate robustness risk. Section \ref{comppp} shows how the approximate STL robustness risk can be  estimated from data, while Section \ref{sec:calc_risk} shows under which conditions it can be computed exactly. The simulation results within CARLA are presented in Section \ref{sec:simulations} followed by conclusions in Section \ref{sec:conclusion}. 

%% file: chapter/background.tex
\section{Background}
\label{sec:backgound}

We first provide background on signal temporal logic, stochastic processes, and risk measures.

\subsection{Signal Temporal Logic}
\label{sec:STL}
Signal temporal logic (STL) is based on deterministic signals $x:T\to\mathbb{R}^n$ where $T$ denotes the time domain \citep{maler2004monitoring}. We particularly consider continuous time $T:=\mathbb{R}$ (the set of real numbers) and discrete time  $T:=\mathbb{Z}$ (the set of natural numbers). The atomic elements of STL are predicates that are functions $\mu:\mathbb{R}^n\to\mathbb{B}$ where  $\mathbb{B}:=\{\top,\bot\}$ is the set of Booleans consisting of the true and false elements $\top:=1$ and $\bot:=-1$, respectively. Let us associate an observation map $O^\mu\subseteq \mathbb{R}^n$ with a predicate  $\mu$ that indicates regions within the state space where the predicate $\mu$ is true, i.e.,
\begin{align*}
O^\mu:=\mu^{-1}(\top)
\end{align*}
where $\mu^{-1}(\top)$ denotes the inverse image of $\top$ under the function $\mu$. We assume throughout the paper that the sets $O^\mu$ and $O^{\neg\mu}$ are non-empty and measurable, which is a mild technical assumption. In other words, the sets $O^\mu$ and $O^{\neg\mu}$  are elements of the Borel $\sigma$-algebra $\mathcal{B}^n$ of $\mathbb{R}^n$.  
\begin{remark}
For convenience, the predicate $\mu$ is often defined via a predicate function $h:\mathbb{R}^n\to\mathbb{R}$ as
\begin{align*}
\mu(\zeta):=\begin{cases}
\top & \text{if } h(\zeta)\ge 0\\
\bot &\text{otherwise}
\end{cases}
\end{align*}
for $\zeta\in\mathbb{R}^n$. In this case, we have  $O^\mu=\{\zeta\in\mathbb{R}^n|h(\zeta)\ge 0\}$.
\end{remark}

The syntax of STL, which recursively allows to formulate system specifications, is defined as 
\begin{align}\label{eq:full_STL}
\phi \; ::= \; \top \; | \; \mu \; | \;  \neg \phi \; | \; \phi' \wedge \phi'' \; | \; \phi'  U_I \phi'' \; | \; \phi' \underline{U}_I \phi'' \,
\end{align}
where $\phi'$ and $\phi''$ are STL formulas and where $U_I$ is the future until operator with time interval $I\subseteq \mathbb{R}_{\ge 0}$, while $\underline{U}_I$ is the past until-operator. The Boolean operators $\neg$ and $\wedge$ encode negations and conjunctions, respectively. We say that an STL formula $\phi$ as in \eqref{eq:full_STL} is bounded if the time interval $I$ is restricted to be compact. Based on these elementary operators, we can define the set of operators
\begin{align*}
\phi' \vee \phi''&:=\neg(\neg\phi' \wedge \neg\phi'') &\text{ (disjunction operator)},\\
F_I\phi&:=\top U_I \phi &\text{ (future eventually operator)},\\
\underline{F}_I\phi&:=\top \underline{U}_I \phi &\text{ (past eventually operator)},\\
G_I\phi&:=\neg F_I\neg \phi &\text{ (future always operator)},\\
\underline{G}_I\phi&:=\neg \underline{F}_I\neg \phi &\text{ (past always operator).}
\end{align*}

\subsubsection{Semantics} To determine whether or not a signal $x:T\to\mathbb{R}^n$ satisfies an STL formula $\phi$, we define the semantics of $\phi$ by means of the satisfaction function $\beta^\phi:\mathfrak{F}(T,\mathbb{R}^n)\times T \to \mathbb{B}$.\footnote{We use the notation $\mathfrak{F}(A,B)$ to denote the set of all measurable functions mapping from the domain $A$ into the domain $B$, i.e.,  an element $f\in \mathfrak{F}(A,B)$ is a measurable function $f:A\to B$.}  In particular, $\beta^\phi(x,t)=\top$ indicates that the signal $x$ satisfies the formula $\phi$ at time $t$, while $\beta^\phi(x,t)=\bot$ indicates that $x$ does not satisfy $\phi$ at time $t$. While the intuitive meanings of the Boolean operators $\neg$ (`not'), $\wedge$ (`and'), and $\vee$ (`or') are clear, we note that the future until operator $\phi' {U}_I \phi''$ encodes that $\phi'$ holds until $\phi''$ holds. Specifically,  $\beta^{\phi' {U}_I \phi''}(x,t)=\top$  means that $\phi'$ holds for all times after $t$ (not necessarily at time $t$) until $\phi''$ holds within the time interval $(t\oplus I)\cap T$.\footnote{We use the notation $\oplus$ and $\ominus$ to denote the Minkowski sum and the Minkowski difference, respectively.} Similarly, $\beta^{F_I \phi}(x,t)=\top$ encodes that $\phi$ holds eventually within $(t\oplus I)\cap T$, while  $\beta^{G_I \phi}(x,t)=\top$ encodes that $\phi$ holds always within $(t\oplus I)\cap T$.   For a formal definition of  $\beta^\phi(x,t)$, we refer to Appendix \ref{app:STL}.

 We are usually interested in the satisfaction function $\beta^\phi(x,0)$  which determines the satisfaction of $\phi$ by $x$ at time zero, the time at which we assume $\phi$ to be enabled.  An STL formula $\phi$ is hence said to be satisfiable if $\exists x\in \mathfrak{F}(T,\mathbb{R}^n)$ such that $\beta^\phi(x,0)=\top$. The following example is taken from \citet{lindemann2021stl} and used as a running example throughout the paper.

\begin{example}\label{ex1}
Consider a delivery robot that needs to  perform two time-critical delivery tasks in regions $A$ and $B$ sequentially while avoiding areas $C$ and $D$, see Fig. \ref{ex:1_figure}. We consider the STL formula 
\begin{align}\label{ex:1_formula}
    \phi:=G_{[0,3]}(\neg \mu_{C} \wedge \neg\mu_{D}) \wedge F_{[1,2]}(\mu_{A} \wedge F_{[0,1]}\mu_{B}).
\end{align}
where the regions $A$, $B$, $C$, and $D$ are encoded by the predicates $\mu_A$, $\mu_B$, $\mu_C$, and $\mu_D$, respectively, that are defined below. Let the state  $x(t)\in\mathbb{R}^{10}$ of the system at time $t$ be
\begin{align*}
    x(t):=\begin{bmatrix}r(t) & a & b& c& d \end{bmatrix}^T
\end{align*} 
where $r(t)$ is the robot position at time $t$ and where $a$, $b$, $c$, and $d$ denote the center points of the regions $A$, $B$, $C$, and $D$ that are defined as
\begin{align*}
    a:=\begin{bmatrix}
        4 & 5
    \end{bmatrix}^T
     \hspace{0.5cm} b:=\begin{bmatrix}
        7 & 2
    \end{bmatrix}^T\hspace{0.5cm}
     c:=\begin{bmatrix}
        2 & 3
    \end{bmatrix}^T
     \hspace{0.5cm}d:=\begin{bmatrix}
        6 & 4
    \end{bmatrix}.^T
\end{align*}  
The predicates $\mu_A$, $\mu_B$, $\mu_C$, and $\mu_D$ are now defined by their observation maps
\begin{align}
    O^{\mu_A}&:=\{x\in\mathbb{R}^{10}|\|r-a\|_\infty\le 0.5\},\nonumber\\
    O^{\mu_B}&:=\{x\in\mathbb{R}^{10}|\|r-b\|_2\le 0.7\},\nonumber\\
    O^{\mu_C}&:=\{x\in\mathbb{R}^{10}|\|r-c\|_\infty\le 0.5\},\label{eq:c_}\\
    O^{\mu_D}&:=\{x\in\mathbb{R}^{10}|\|r-d\|_2\le 0.7\}\label{eq:d_}.
\end{align}
where $\|\cdot\|_2$ is the Euclidean and $\|\cdot\|_\infty$ is the infinity norm. In Fig. \ref{ex:1_figure}, six different robot trajectories $r_1$-$r_6$ are shown. It can be seen that the signal $x_1$ that corresponds to $r_1$ violates $\phi$, while $x_2$-$x_6$ satisfy $\phi$, i.e., we have $\beta^\phi(x_1,0)=\bot$ and $\beta^\phi(x_j,0)=\top$ for all $j\in\{2,\hdots,6\}$.
\begin{figure}
\centering
\includegraphics[scale=0.5]{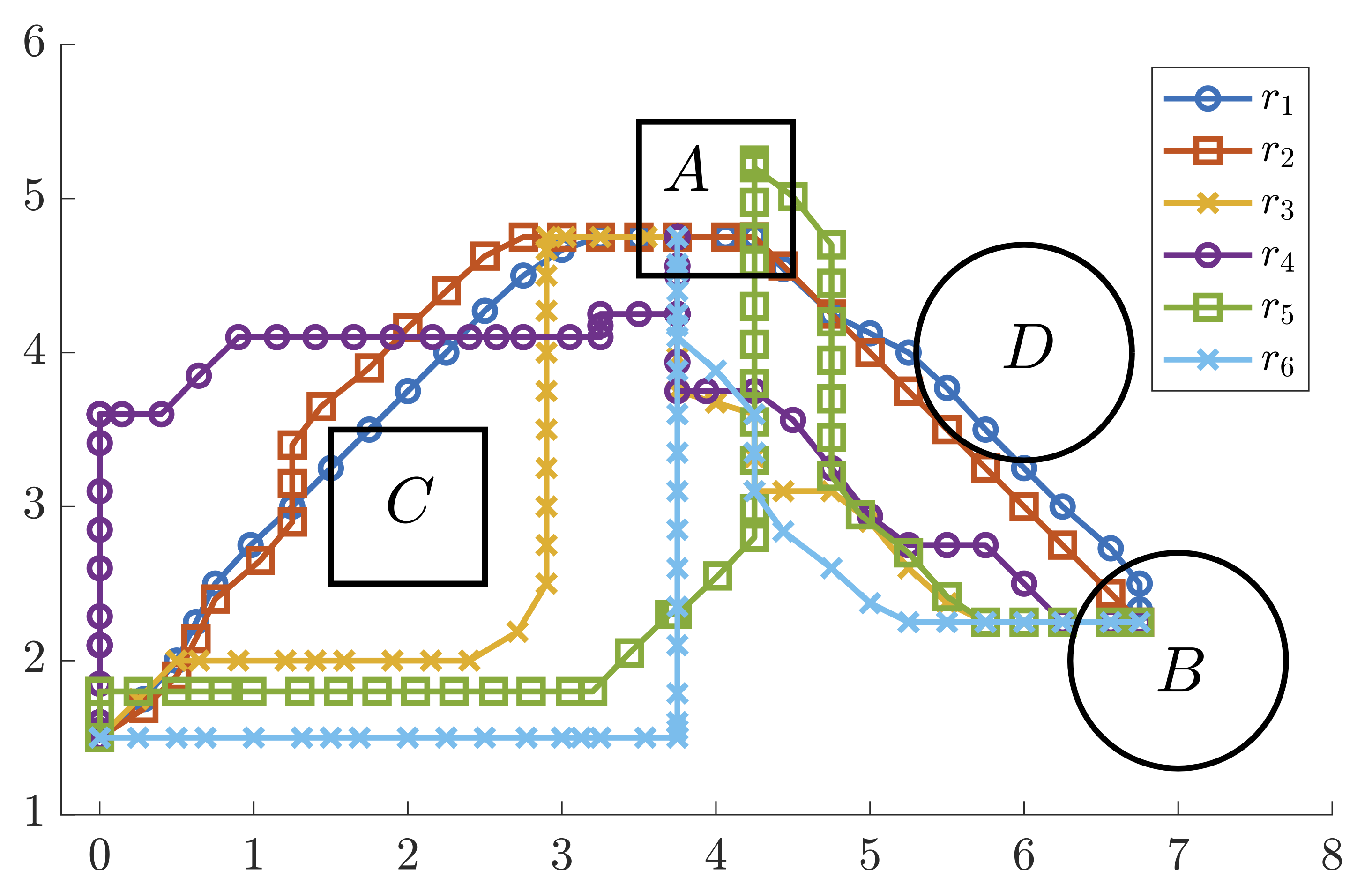}
\caption{The figure shows six potential robot trajectories $r_1$-$r_6$ and the four regions $A$, $B$, $C$, and $D$. The specification given in \eqref{ex:1_formula} is violated by $r_1$ and satisfied by $r_2$-$r_6$. It can be seen that $r_2$ only marginally satisfies  $\phi$, while $r_3$-$r_6$ satisfy $\phi$ robustly.}
\label{ex:1_figure}
\end{figure}
\end{example}

\begin{remark}\label{remmm}
The operators ${U}_I$ and  $\underline{U}_I$ are  the strict non-matching versions of the until operators. In particular, $\phi' {U}_I \phi''$ is: 1) strict as it does not require $\phi'$ to hold at the current time $t$, and 2) non-matching as it does not require that $\phi'$ and $\phi''$ have to hold at the same time. When dealing with continuous-time stochastic systems later in this paper, we replace the  strict non-matching versions ${U}_I$ and $\underline{U}_I$ by the non-strict  matching versions that we denote by $\vec{U}_I$ and $\vec{\underline{U}}_I$, see Appendix \ref{app:STL} for their formal definitions. We note that STL with until operators ${U}_I$ and $\underline{U}_I$ is more expressive than STL with $\vec{U}_I$ and $\vec{\underline{U}}_I$. When excluding Zeno-signals, there is however no difference between these two notions \citep{furia2007expressiveness}. As one rarely encounters Zeno-signals, we argue that the restriction to the non-strict matching version of the until operator for continuous-time stochastic processes is not restrictive in practice.
\end{remark}

\subsubsection{Robustness Degree}	Importantly, one may also be interested in the quality of satisfaction and additionally ask how robustly the signal $x$ satisfies the STL formula $\phi$ at time $t$. To answer this question, the authors in \citet[Definition 7]{fainekos2009robustness} define the \emph{robustness degree} that we recal next in a slightly modified manner. If $\beta^\phi(x,t)=\top$, the robustness degree  quantifies how much the signal $x$ can be perturbed by additive noise before changing the value of $\beta^\phi(x,t)$. Towards a formal definition, let us first define the set of  signals that \emph{violate} $\phi$ at time $t$ as
	\begin{align*}
	\mathcal{L}^{\neg\phi}(t):=\{x\in \mathfrak{F}(T,\mathbb{R}^n)|\beta^{\neg\phi}(x,t)=\top\}.
	\end{align*} 
To measure  distances between signals, let us define the  metric $\kappa:\mathfrak{F}(T,\mathbb{R}^n)\times\mathfrak{F}(T,\mathbb{R}^n)\to\overline{\mathbb{R}}_{\ge 0}$ as
	\begin{align*}
	\kappa(x,x'):=\sup_{t\in T} d\big(x(t),x'(t)\big)
	\end{align*}
	where $\overline{\mathbb{R}}_{\ge 0}:=\mathbb{R}_{\ge 0}\cup\{\infty\}$ is the set of nonnegative extended real numbers and where $d:\mathbb{R}^n\times\mathbb{R}^n\to\mathbb\overline{\mathbb{R}}_{\ge 0}$ is a metric assigning a distance in $\mathbb{R}^n$, e.g., the Euclidean norm. Throughout the paper, we use the extended definitions of the supremum and infimum operators, e.g., $\sup \mathbb{R}= \infty$. Note that $\kappa(x,x')$ is the $L_\infty$ norm of the signal $x-x'$ and measures the  distance between the signals $x$ and $x'$.

To set some general notation, for a metric space $(S,\kappa)$ with metric $\kappa$ we denote by 
\begin{align*}
    \bar{\kappa}(x,S'):=\inf_{x'\in S'} \kappa(x,x')
\end{align*} the distance of a point $x\in S$ to a nonempty set $S'\subseteq S$. Using this definition, the robustness degree $\text{RD}^\phi:\mathfrak{F}(T,\mathbb{R}^n)\times T\to \overline{\mathbb{R}}_{\ge 0}$ is now defined  via the metric $\kappa$ as the distance of the signal $x$ to the set of violating signals $\mathcal{L}^{\neg\phi}(t)$.
	\begin{definition}[Robustness Degree\footnote{The robustness degree in \citet[Definition~7]{fainekos2009robustness} is defined slightly differently by instead considering the signed distance of the signal $x$ to the set of violating signals $\mathcal{L}^{\neg\phi}(t)$.}]\label{def:rd_cont}
	For a signal $x:T\to\mathbb{R}^n$ and an STL formula $\phi$, the robustness degree $\text{RD}^{\phi}(x,t)$ is defined as
	\begin{align*}
	\text{RD}^{\phi}(x,t):=\bar{\kappa}\big(x,\text{cl}(\mathcal{L}^{\neg\phi}(t))\big)
	\end{align*} 
	where $\text{cl}(\mathcal{L}^{\neg\phi}(t))$ denotes the closure of the set $\mathcal{L}^{\neg\phi}(t)$.
	\end{definition}

 By definition of the robustness degree, the following properties hold. If $\text{RD}^{\phi}(x,t)> 0$, then $\beta^\phi(x,t)=\top$, i.e., the signal $x$ satisfies $\phi$ at time $t$. It further follows that all signals $x'\in\mathfrak{F}(T,\mathbb{R}^n)$ with $\kappa(x,x')<\text{RD}^{\phi}(x,t)$ are such that $\beta^\phi(x',t)=\top$.  The robustness degree defines in fact a \emph{robust neighborhood}, which is a set strictly containing $x$, so that for all $x'$ in this robust neighborhood we have $\beta^\phi(x,t)=\beta^\phi(x',t)$. Finally note that $\text{RD}^{\phi}(x,t)=0$ may imply either $\beta^\phi(x,t)=\top$ or $\beta^\phi(x,t)=\bot$, i.e., the signal $x$ either satisfies or violates $\phi$ at time $t$.

\subsubsection{Robust Semantics} Note that it is in general difficult to calculate the robustness degree $\text{RD}^{\phi}(x,t)$ as the set $\mathcal{L}^{\neg\phi}(t)$ is hard to calculate. The  authors in \citet{fainekos2009robustness} introduce the \emph{robust semantics} $\rho^\phi:\mathfrak{F}(T,\mathbb{R}^n)\times T\to \overline{\mathbb{R}}$ as an alternative way of finding a robust neighborhood where $\overline{\mathbb{R}}:=\mathbb{R}\cup\{-\infty,\infty\}$ is, in direct analogy to $\overline{\mathbb{R}}_{\ge 0}$, the set of extended real numbers.
\begin{definition}[Robust Semantics]\label{def:quantitative_semantics}
For a signal $x:T\to\mathbb{R}^n$ and an STL formula $\phi$, the robust semantics $\rho^\phi(x,t)$   are recursively defined as 
	\begin{align*}
	\rho^{\top}(x,t)& := \infty,\\
	\rho^{\mu}(x,t)& := \begin{cases} \bar{d}\big(x(t),\text{cl}(O^{\neg\mu})\big) &\text{if } x(t)\in O^{\mu}\\
	-\bar{d}\big(x(t),\text{cl}(O^{\mu})\big) &\text{otherwise,}
	\end{cases}\\
	\rho^{\neg\phi}(x,t) &:= 	-\rho^{\phi}(x,t),\\
	\rho^{\phi' \wedge \phi''}(x,t) &:= 	\min(\rho^{\phi'}(x,t),\rho^{\phi''}(x,t)),\\
	%	\rho^{\phi' \vee \phi''}(x,t) &:= 	\max(\rho^{\phi'}(x,t),\rho^{\phi''}(x,t)),\\
	\rho^{\phi' U_I \phi''}(x,t) &:= \underset{t''\in (t\oplus I)\cap T}{\text{sup}}  \Big(\min\big(\rho^{\phi''}(x,t''),\underset{t'\in (t,t'')\cap T}{\text{inf}}\rho^{\phi'}(x,t') \big)\Big), \\
	\rho^{\phi' \underline{U}_I \phi''}(x,t) &:= \underset{t''\in (t\ominus I)\cap T}{\text{sup}} \Big( \min\big(\rho^{\phi''}(x,t''),  \underset{t'\in (t'',t)\cap T}{\text{inf}}\rho^{\phi'}(x,t') \big)\Big).
	%	\rho^{G_I \phi}(x,t) &:= \underset{t'\in t\oplus I}{\text{inf}}\rho^{\phi}(x,t'),\\
	%	\rho^{\underline{G}_I \phi}(x,t) &:= \underset{t'\in t\ominus I}{\text{inf}}\rho^{\phi}(x,t'),\\
	%	\rho^{F_I \phi}(x,t) &:= \underset{t'\in t\oplus I}{\text{sup}}\rho^{\phi}(x,t'),\\
	%	\rho^{\underline{F}_I \phi}(x,t) &:= \underset{t'\in t\ominus I}{\text{sup}}\rho^{\phi}(x,t').
	\end{align*}
\end{definition}

\begin{remark}
With respect to Remark \ref{remmm}, the non-strict matching version of the until operators replace the open time intervals $(t,t'')$ in Definition \ref{def:quantitative_semantics} by the closed time intervals $[t,t'']$ so that
	\begin{align*}
	\rho^{\phi' \vec{U}_I \phi''}(x,t) &:= \underset{t''\in (t\oplus I)\cap T}{\text{sup}}  \Big(\min\big(\rho^{\phi''}(x,t''),\underset{t'\in [t,t'']\cap T}{\text{inf}}\rho^{\phi'}(x,t') \big)\Big), \\
	\rho^{\phi' \vec{\underline{U}}_I \phi''}(x,t) &:= \underset{t''\in (t\ominus I)\cap T}{\text{sup}} \Big( \min\big(\rho^{\phi''}(x,t''),  \underset{t'\in [t'',t]\cap T}{\text{inf}}\rho^{\phi'}(x,t') \big)\Big).
	\end{align*}
\end{remark}

%Finally, we provide the following remark concerning the definition of the robust semantics of predicates $\rho^{\mu}(x,t)$.
%\begin{remark}\label{rem:222}
%Recall that $\text{dist}^{\mu}(x,t)=\bar{\kappa}\big(x,\text{cl}(\mathcal{L}^\mu(t))\big)$. However, as shown in \citet[Lemma 57]{fainekos2009robustness}, it holds that
%	\begin{align*}
%	\text{dist}^{\mu}(x,t)= \bar{d}\big(x(t),\text{cl}(O^\mu)\big):=\inf_{x'\in \text{cl}(O^\mu)} d(x(t),x').
%	\end{align*}  Consequently, $\rho^{\mu}(x,t)$ encodes the signed distance from the signal $x(t)$ to the set $ O^\mu$. It also holds that $x\in\mathcal{L}^\mu(t)$ if and only if $x(t)\in O^\mu$.
%\end{remark}
	
Importantly, by slight modification of \citet[Theorem 28]{fainekos2009robustness}, we know that 
\begin{align}\label{underapprox}
 \rho^{\phi}(x,t)\le \text{RD}^{\phi}(x,t).
\end{align} 
The robust semantics $\rho^\phi(x,t)$ hence provides a tractable under-approximation of the robustness degree $\text{RD}^{\phi}(x,t)$. The robust semantics are sound in the  sense that $\beta^\phi(x,t)=\top$  if  $\rho^\phi(x,t)> 0$ and $\beta^\phi(x,t)=\bot$ if $\rho^\phi(x,t)< 0$ \citep[Proposition 30]{fainekos2009robustness}.
\begin{myexpcont}
Consider again the trajectories shown in Fig. \ref{ex:1_figure}. We obtain $\rho^\phi(x_1,0)=-0.15$, $\rho^\phi(x_2,0)=0.01$, and $\rho^\phi(x_j,0)=0.25$ for all $j\in\{3,\hdots,6\}$. The reason for $x_1$ having negative robustness lies in $r_1$ intersecting with the region $D$. Marginal robustness of $x_2$ is explained as $r_2$ only marginally avoids the region $D$ while all other trajectories avoid the region $D$ robustly.
\end{myexpcont}

\subsection{Random Variables and Stochastic Processes}
\label{sec:stoch}
Instead of interpreting an STL specifications $\phi$ over deterministic signals, we will interpret $\phi$ over stochastic processes. Consider therefore the \emph{probability space} $(\Omega,\mathcal{F},P)$  where $\Omega$ is the sample space, $\mathcal{F}$ is a $\sigma$-algebra of $\Omega$, and $P:\mathcal{F}\to[0,1]$ is a probability measure.

 Let $Z$ denote a real-valued \emph{random vector}, i.e., a measurable function $Z:\Omega\to\mathbb{R}^n$. When $n=1$, we  say $Z$ is a \emph{random variable}. We refer to $Z(\omega)$ as a realization of the random vector $Z$ where $\omega\in\Omega$.   Since $Z$ is a measurable function, a probability space can be defined for $Z$ so that probabilities can be assigned to events related to values of $Z$.\footnote{Particularly, this probability space is $(\mathbb{R}^n,\mathcal{B}^n,P_Z)$ where, for Borel sets $B\in\mathcal{B}^n$, the probability measure $P_Z:\mathcal{B}^n\to[0,1]$ is defined as 
$P_Z(B):=P(Z^{-1}(B))$
where $Z^{-1}(B):=\{\omega\in\Omega|Z(w)\in B\}$ is the inverse image of $B$ under $Z$.} Consequently, a cumulative distribution function (CDF) $F_Z(z)$ can be defined for $Z$. Given a random vector $Z$, we can derive other random variables. Assume for instance a measurable function $g: \mathbb{R}^n \to \mathbb{R}$, then $g(Z(\omega))$ becomes a \emph{derived random variable} since function composition preserves measureability, see e.g.,  \citet{durrett2019probability} for more details.

A \emph{stochastic process} is a function $X:T\times \Omega \to \mathbb{R}^n$ where  $X(t,\cdot)$ is a random vector for each fixed $t\in T$.  A stochastic process can   be viewed as a collection of random vectors $\{X(t,\cdot)|t\in T\}$  that are defined on a common probability space $(\Omega,\mathcal{F},P)$ and that are indexed by $T$.  For a fixed $\omega\in\Omega$, the function $X(\cdot,\omega)$ is a \emph{realization} of the stochastic process.  Another interpretation is that a stochastic process is a collection of deterministic functions of time $\{X(\cdot,\omega)|\omega\in \Omega\}$ that are indexed by $\Omega$. 
 
 %We refer to $X$ as continuous-space stochastic process when the CDF of $X(t,\cdot)$ is continuous and as discrete-space stochastic process otherwise.  

\subsection{Risk Measures}
\label{sec:risk}

\label{sec:risk_measures}
	\begin{figure}
	\centering
	\includegraphics[scale=0.4]{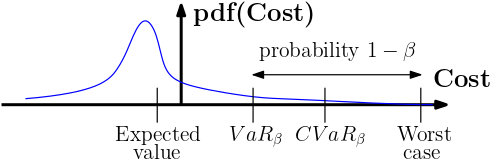}
	\caption{Illustration of the expected value, the value-at-risk, and the conditional value-at-risk.}
	\label{ex:2_figure}
\end{figure}

A \emph{risk measure} is a function $R:\mathfrak{F}(\Omega,\mathbb{R})\to \mathbb{R}$ that maps from the set of real-valued random variables to the real numbers. In particular, we refer to the input of a risk measure $R$ as the \emph{cost random variable} since typically a cost is associated with the input of $R$. Risk measures hence allow for a risk assessment in terms of such cost random variables.

In this paper, we particularly use the expected value, the value-at-risk $VaR_\beta$, and the conditional value-at-risk $CVaR_\beta$ at risk level $\beta\in (0,1)$ which are commonly used risk measures, see Figure \ref{ex:2_figure}. The $VaR_\beta$ of a  random variable $Z:\Omega\to\mathbb{R}$ is defined as
\begin{align*}
VaR_\beta(Z):=\inf \{ \alpha \in \mathbb{R} |  F_Z(\alpha) \ge \beta \}, 
\end{align*}
i.e., the right $1-\beta$  quantile of $Z$. The $CVaR_\beta$ of $Z$ is defined as
\begin{align*}
CVaR_\beta(Z):=\inf_{\alpha \in \mathbb{R}}  \big(\alpha+(1-\beta)^{-1}E([Z-\alpha]^+)\big)
\end{align*} 
where $[Z-\alpha]^+:=\max(Z-\alpha,0)$. When the CDF $F_Z$ of $Z$ is continuous, it holds that $CVaR_\beta(Z):=E(Z|Z\ge VaR_\beta(Z))$, i.e., $CVaR_\beta(Z)$ is the expected value of $Z$ conditioned on the events where $Z$ is greater or equal than  $VaR_\beta(Z)$.

There are various desriable properties that a risk measure $R$ may satisfy, see  \citet{majumdar2020should} for more information. We  emphasize that our presented method is compatible with any monotone  risk measure, where monotonicity of $R$ is  defined as follows:
\begin{itemize}
	\item For two cost random variables $Z,Z'\in \mathfrak{F}(\Omega,\mathbb{R})$, the risk measure $R$ is \emph{monotone} if 
	\begin{align*}
	Z(\omega) \leq Z'(\omega) \text{ for all } \omega\in\Omega \;\; \implies \;\; R(Z) \le R(Z').
	\end{align*}
\end{itemize} 

The assumption of considering monotone risk measures is very mild, and both the value-at-risk $VaR_\beta(Z)$ and the conditional value-at-risk $CVaR_\beta(Z)$ as well as the expected value are monotone.

%% file: chapter/problem_formulation.tex
\section{The Risk of Lacking Robustness against Failure}
\label{risskk}

We interpret STL formulas $\phi$ over stochastic processes $X$ instead of deterministic signals $x$. It is, however, not immediately clear how to interpret the satisfaction of $\phi$ by $X$. One way is to argue about the probability of satisfaction, see e.g., \citet{farahani2018shrinking}, but probabilities provide no information about the risk and the robustness of $X$ with respect to $\phi$. In fact, some realizations of $X$ may satisfy $\phi$ robustly, while some other realizations of $X$ may  satisfy  $\phi$ only marginally or even  violate $\phi$. This observation leads us to the use of risk measures $R$ to be able to argue about the risk of the stochastic process $X$ lacking robustness against failure of the specification~$\phi$.

%For  ease of notation, let us equivalently refer to the stochastic process of $X$ as the function $X_t:\Omega\to\mathfrak{F}(T,\mathbb{R}^n)$ with
%\begin{align*}
%X_t:=X(\cdot,\omega)
%\end{align*}  
%so that $X_t(\omega')(t')=X(t',\omega')$ for $t'\in T$ and $\omega'\in\Omega$. 

%% file: chapter/main.tex
\subsection{Measurability of Semantics,  Robustness Degree, and Robust Semantics}

To define the risk of a stochastic process $X$, we first need to show under which conditions the semantics $\beta^\phi(X,t)$, the robustness degree $\text{RD}^{\phi}(X,t)$ , and the robust semantics  $\rho^\phi(X,t)$ are derived random variables.  For discrete-time stochastic processes, no  assumptions have to be made. 
\begin{theorem}\label{thm:1}
	Let $X$ be a discrete-time stochastic process, i.e., $T:=\mathbb{Z}$. Let $\phi$ be an STL specification as in \eqref{eq:full_STL}. Then $\beta^\phi(X(\cdot,\omega),t)$, $\text{RD}^\phi(X(\cdot,\omega),t)$,  and $\rho^\phi(X(\cdot,\omega),t)$  are measurable in $\omega$ for a fixed $t\in T$, i.e.,  $\beta^\phi(X,t)$, $\text{RD}^\phi(X,t)$, and $\rho^\phi(X,t)$ are random variables.
\end{theorem}

For continuous-time stochastic processes, however, we have to impose additional technical assumptions. Particularly, we have to  restrict the class of STL formulas in \eqref{eq:full_STL} and make further assumptions on the stochastic process $X$.

\begin{theorem}\label{thm:2}
	Let $X$ be a continuous-time stochastic process, i.e., $T:=\mathbb{R}$. Let  $\phi$ be a bounded STL specification as in \eqref{eq:full_STL}, but where the strict non-matching until operators  ${U}_I$ and ${\underline{U}}_I$ are  replaced with the non-strict matching until operators $\vec{U}_I$ and $\vec{\underline{U}}_I$. Then $\beta^\phi(X(\cdot,\omega),t)$ is measurable in $\omega$ for a fixed $t\in T$, i.e.,  $\beta^\phi(X,t)$ is a random variable. If $X(\cdot,\omega):\Omega\to\mathfrak{F}(T,\mathbb{R}^n)$ is measureable\footnote{Here, we mean measurable with respect to the Borel $\sigma$-algebras induced by the Skorokhod metric, see \citep{bartocci2015system} for details.}, then $\text{RD}^\phi(X(\cdot,\omega),t)$ is measurable in $\omega$ for a fixed $t\in T$, i.e., $\text{RD}^\phi(X,t)$ is a random variable, and if additionally $X(\cdot,\omega)$ is a cadlag function\footnote{Cadlag functions are right continuous functions with left limits.} for each $\omega\in\Omega$, then $\rho^\phi(X(\cdot,\omega),t)$ is measurable in $\omega$ for a fixed $t\in T$, i.e.,  $\rho^\phi(X,t)$ is a random variable.\footnote{The result for measurability of $\rho^\phi(X(\cdot,\omega),t)$ is mainly taken from  \citep[Theorem 6]{bartocci2015system}.}
\end{theorem}

Consequently, the probabilities
$P(\beta^{\phi}(X,t)\in B)$,  $P(\rho^{\phi}(X,t)\in B)$, and $P(\text{RD}^{\phi}(X,t)\in B)$\footnote{We use the shorthand notations $P(\beta^{\phi}(X,t)\in B)$, $P(\rho^{\phi}(X,t)\in B)$, and $P(\text{RD}^{\phi}(X,t)\in B)$ instead of  $P(\{\omega\in\Omega|\beta^{\phi}(X(\cdot,\omega),t)\in B\})$, $P(\{\omega\in\Omega|\rho^{\phi}(X(\cdot,\omega),t)\in B\})$, and $P(\{\omega\in\Omega|\text{RD}^{\phi}(X(\cdot,\omega),t)\in B\})$, respectively.} are well defined for measurable sets $B$ from the corresponding measurable spaces. This enables us to define the STL robustness risk in the next section.

\begin{remark}\label{rem:comparison}
We first note that the assumption of a bounded STL formula $\phi$ with the non-strict matching until operator is made for a technical reason. While the restriction to bounded formulas limits our expressivity to finite time specifications, the consideration of the non-strict matching until operator is not restrictive as discussed in Remark \ref{remmm}. We remark  that \citet{bartocci2015system} showed measurability of $\rho^\phi(X(\cdot,\omega),t)$ under the assumption of a bounded STL specification $\phi$ with non-strict matching until operators, while we additionally show measurability of the semantics $\beta^\phi(X(\cdot,\omega),t)$ and the robustness degree $\text{RD}^\phi(X(\cdot,\omega),t)$ without any additional continuity assumptions on $X$.  Lastly, we recall that we do not need to assume that $\phi$ is bounded for a discrete-time stochastic process as per Theorem  \ref{thm:1}. 
\end{remark}

\subsection{The STL Robustness Risk}

One way of defining the risk associated with a stochastic process $X$ is to consider the satisfaction function $\beta^\phi(X,t)$. However, not much information about the robustness of $X$ can be inferred due to binary encoding of  $\beta^\phi(X,t)$. Instead, we consider the risk of the stochastic process $X$ lacking robustness against failure of the specification $\phi$  by considering the robustness degree $\text{RD}^{\phi}(X,t)$. 
 \begin{example}
 \begin{figure*}
	\centering
	\includegraphics[scale=0.19]{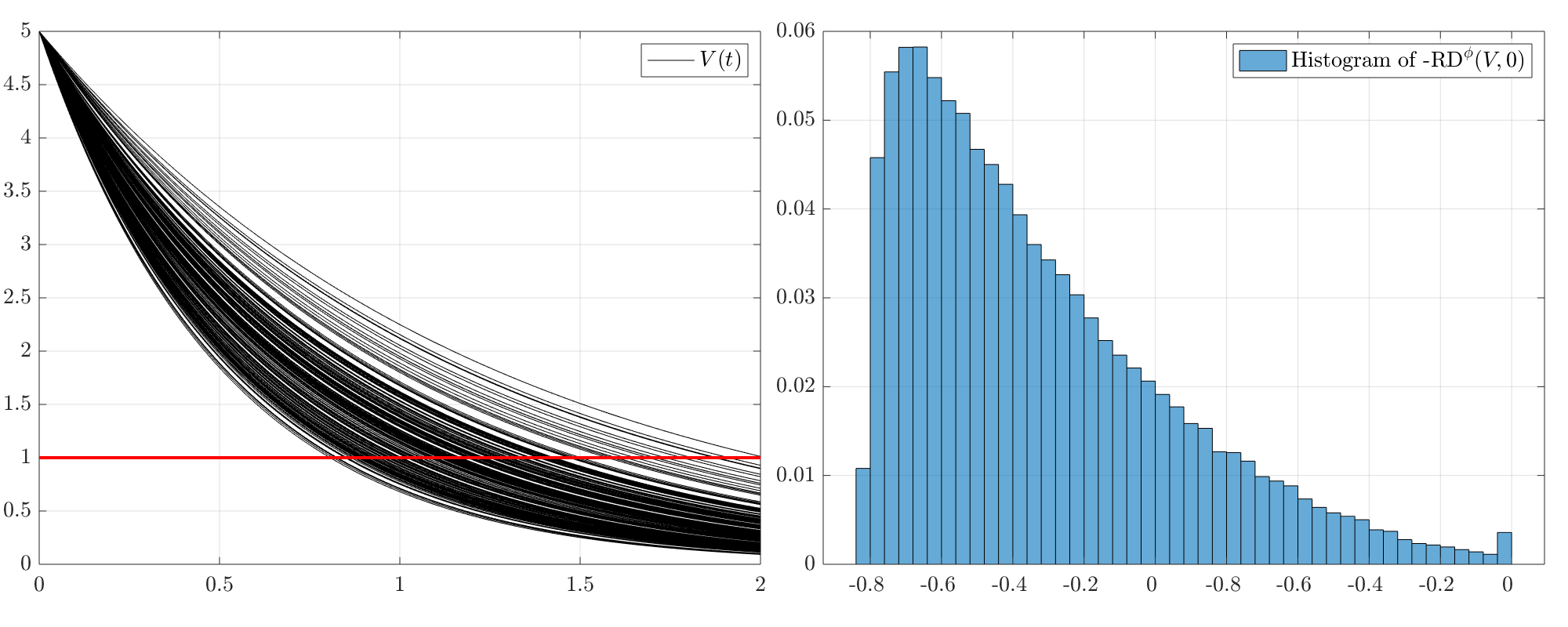}
	\caption{Left: $200$ realizations of the voltage $V(t)$ over the capacitor of an RC circuit. Right: Histogram of the negative robustness degree $-\text{RD}^{\phi}(V,0)$ of the specification $\phi:=G_{[2,\infty)}(V\le 1)$.}
	\label{fig:RC}
\end{figure*}
Consider an electric RC circuit consisting of a resistor with resistance $\mathcal{R}$ and a capacitor with capacitance $\mathcal{C}:=1$. If the capacitor is initially charged with $V_0:=5$, then the capacitor discharges its  energy over time once the circuit is closed. In fact, the voltage over the capacitor is described by 
\begin{align*}
    V(t)=V_0\exp(-\tau t)
\end{align*}
where $\tau:=1/\mathcal{R}\mathcal{C}$ is the time constant. Assume that the resistance is unknown and modeled as $\mathcal{R}:=0.5+Z$ where $Z$ is a random variable following a beta distribution with  probability density function $f_Z(z):=\frac{1}{B(1.5,5)} z^{1.5-1} (1-z)^{5-1}$ where $B(1.5,5)$ is the beta function with parameters $1.5$ and $5$. Consequently, the voltage $V$ becomes a stochastic process of which we plot $200$ realizations in Fig. \ref{fig:RC} (left). As a specification $\phi$, we want that the voltage $V(t)$ drops below $1$ after $2$ s, i.e., 
\begin{align*}
    \phi:=G_{[2,\infty)}(V\le 1).
\end{align*}
In Fig. \ref{fig:RC} (right), we show the histogram of the negative robustness degree $-\text{RD}^{\phi}(V,0)$ for $100000$ realizations. To estimate the risk of the stochastic process $X$ lacking robustness against failure of $\phi$, we can now compose $-\text{RD}^{\phi}(V,0)$ with a risk measure $R$. For instance, the value-at-risk at level $\beta:=0.9$ is $VaR_{0.9}(-\text{RD}^{\phi}(V,0))\approx -0.38$. Recall that $VaR_{0.9}(-\text{RD}^{\phi}(V,0))$ is the $0.1$ quantile of $-\text{RD}^{\phi}(V,0)$. This means that with a probability of at least $0.9$ the robustness degree is not smaller (i.e., greater) than $|VaR_{0.9}(-\text{RD}^{\phi}(V,0))|\approx 0.38$ or, in other words, that in at most $10$ percent of the cases the robustness is smaller than $0.38$. This information is useful as it allows us to quantify how much uncertainty our system can handle, e.g., when we do not know the value of $V_0$ exactly.
 \end{example}

The previous example motivates the following definition for the risk of the stochastic process $X$ lacking robustness against failure of $\phi$  to which we refer as the STL robustness risk for brevity.
 \begin{definition}[STL Robustness Risk]\label{def:rr}
 	Given an STL formula $\phi$ and a stochastic process $X:T\times\Omega\to\mathbb{R}^n$, the risk of $X$ lacking robustness against failure of $\phi$  at time $t$ is defined as 
 	\begin{align*}
 	R(-\text{RD}^{\phi}(X,t)).
 	\end{align*}  
 \end{definition}

We  remark that a large positive value of $\text{RD}^{\phi}(X(\cdot,\omega),t)$ for a realization $\omega\in\Omega$ indicates robust satisfaction of $\phi$. Therefore, the negative robustness degree $-\text{RD}^{\phi}(X,t)$ is the cost random variable that is chosen as the input for the risk measure $R$.  This way, a large robustness degree results in a low cost.  Finally, note that $R(-\text{RD}^\phi(X',t))\le R(-\text{RD}^\phi(X'',t))$ implies that the stochastic process $X'$ is less risky than the stochastic process $X''$ with respect to the specification $\phi$.

\subsection{The Approximate STL Robustness Risk} 
Unfortunately, the STL robustness risk $R(-\text{RD}^{\phi}(X,t))$ can in general not be calculated as the robustness degree in Definition \ref{def:rd_cont} is  difficult to calculate. Instead, we will focus on $R(-\rho^\phi(X,t))$ using the robust semantics as an approximation of the STL robustness risk. 
 \begin{definition}[Approximate STL Robustness Risk]\label{def:rrr}
 	Given an STL formula $\phi$ and a stochastic process $X:T\times\Omega\to\mathbb{R}^n$, the \emph{approximate} risk of $X$ lacking robustness against failure of $\phi$ at time $t$ is defined as 
 	\begin{align*}
 	R(-\rho^\phi(X,t)).
 	\end{align*}  
 \end{definition}

Fortunately, the approximate STL robustness risk $R(-\rho^\phi(X,t))$  over-approximates the STL robustness risk $R(-\text{RD}^{\phi}(X,t))$ when $R$ is a monotone risk measure as shown next. 
\begin{theorem}\label{thm:3}
	Let $X$ be a stochastic process, $\phi$ be an STL specification as in \eqref{eq:full_STL}, and $R$ be a monotone risk measure. Then it holds that 
	\begin{align*}
	    R(-\text{RD}^{\phi}(X,t))\le R(-\rho^\phi(X,t)).
	\end{align*}
\end{theorem}

The previous result is important as using $R(-\rho^\phi(X,t))$ instead of $R(-\text{RD}^{\phi}(X,t))$ will not result in an optimistic risk assessment. Especially in safety critical applications, it is desirable to be more risk-averse as opposed to being overly optimistic.

Sometimes one may be interested in scaling the robustness degree to associate a monetary cost with  $\text{RD}^{\phi}(X,t)$ to reflect the severity of events with low robustness. Let us for this purpose consider an increasing cost function $C:\mathbb{R}\to\mathbb{R}$.
\begin{corollary}
	Let $X$ be a stochastic process, $\phi$ be an STL specification as in \eqref{eq:full_STL}, $R$ be a monotone risk measure, and $C$ be an increasing cost function. Then it holds that 
	\begin{align*}
	    R(C(-\text{RD}^{\phi}(X,t)))\le R(C(-\rho^\phi(X,t))).
	\end{align*}
\end{corollary}

\section{Data-Driven Estimation of the Approximate STL  Robustness Risk}
\label{comppp}

In this section, we show how the approximate STL robustness risk $R(-\rho^\phi(X,t))$ can be estimated from data. We assume that we have observed  $N$ independent realizations of the stochastic process $X$, i.e., we know $N$ realizations $X(\cdot,\omega^1),\hdots,X(\cdot,\omega^N)$ where  $\omega^1,\hdots,\omega^N\in\Omega$ are drawn independently and according to the probability measure $P$. A practical example  would be a simulator from which we can unroll trajectories $X(\cdot,\omega^i)$.  For brevity, we denote $X(\cdot,\omega^1),\hdots,X(\cdot,\omega^N)$ by $X^1,\hdots,X^N$. In this way, one can think of $X^1,\hdots,X^N$ as $N$ independent copies of $X$.  We emphasize that we do not need knowledge of the distribution of $X$. Our goal is to derive upper bounds of $R(-\rho^\phi(X,t))$ that hold with high probability. Let us, for convenience, first define the random variable 
\begin{align*}
    Z:=-\rho^\phi(X,t).
\end{align*}
For further convenience, let $Z^i:=-\rho^\phi(X^i,t)$ and let us also define the tuple
\begin{align*}
    \mathcal{Z}:=(Z^1,\hdots,Z^N).
\end{align*}

We consider the value-at-risk $VaR_\beta(Z)$, the conditional value-at-risk $CVaR_\beta(Z)$, and the mean $E(Z)$. Particularly, we derive upper bounds $\overline{VaR}_\beta(\mathcal{Z},\delta)$, $\overline{CVaR}_\beta(\mathcal{Z},\delta)$, and $\overline{E}(\mathcal{Z},\delta)$ that hold with a probability of at least $1-\delta$. By Theorem \ref{thm:3} and Propositions \ref{thm:mmm_}, \ref{prop:2}, and \ref{thm:mmmmm} (presented in the remainder), we then have computational algorithms to find tight upper bounds for the approximate STL  robustness risk and hence for the STL robustness risk, and it hold that with a probability of $1-\delta$
\begin{align*}
    &VaR_\beta(-\text{RD}^{\phi}(X,t))\le VaR_\beta(Z)\le \overline{VaR}_\beta(\mathcal{Z},\delta),\\
    &CVaR_\beta(-\text{RD}^{\phi}(X,t))\le CVaR_\beta(Z)\le \overline{CVaR}_\beta(\mathcal{Z},\delta),\\
    &E(-\text{RD}^{\phi}(X,t))\le E(Z)\le \overline{E}(\mathcal{Z},\delta).
\end{align*}

\subsection{Value-at-Risk (VaR)}  For a risk level of $\beta\in(0,1)$, recall that the VaR of $Z$ is given by
\begin{align*}
VaR_\beta(Z):= \inf\{\alpha\in\mathbb{R}|F_{Z}(\alpha)\ge \beta\}
\end{align*}
where  $F_{Z}(\alpha)$ denotes the CDF of $Z$. To estimate $F_{Z}(\alpha)$, we define the empirical CDF as
\begin{align*}
\widehat{F}(\alpha,\mathcal{Z}):=\frac{1}{N}\sum_{i=1}^N \mathbb{I}(Z^i\le \alpha)
\end{align*}
where $\mathbb{I}$ denotes the indicator function defined as
\begin{align*}
\mathbb{I}(Z^i\le \alpha):=\begin{cases}
1 &\text{if } Z^i\le \alpha\\
0 &\text{otherwise}.
\end{cases}
\end{align*} 
Let now $\delta\in(0,1)$ be a probability threshold. Inspired by \citet{szorenyi2015qualitative}, we calculate an upper bound of $VaR_\beta(Z)$ as
\begin{align*}
\overline{VaR}_\beta(\mathcal{Z},\delta)&:=\inf\Big\{\alpha\in {\mathbb{R}}|\widehat{F}(\alpha,\mathcal{Z})-\sqrt{\frac{\ln(2/\delta)}{2N}}\ge \beta\Big\}
\end{align*}
and a lower bound as
\begin{align*}
\underline{VaR}_\beta(\mathcal{Z},\delta)&:=\inf\Big\{\alpha\in {\mathbb{R}}|\widehat{F}(\alpha,\mathcal{Z})+\sqrt{\frac{\ln(2/\delta)}{2N}}\ge \beta\Big\}
\end{align*}
where we recall that $\inf \emptyset=\infty$ for $\emptyset$ being the empty set due to  the extended definition of the infimum operator. We next show that $\overline{VaR}_\beta(\mathcal{Z},\delta)$ and $\underline{VaR}_\beta(\mathcal{Z},\delta)$ are upper and lower bounds of $VaR_\beta(Z)$, respectively, with a probability of at least $1-\delta$.

\begin{proposition}\label{thm:mmm_}
    Assume that $F_Z$ is  continuous  and let $\delta\in(0,1)$ be a probability threshold and $\beta\in(0,1)$ be a risk level. Let $\overline{VaR}_\beta(\mathcal{Z},\delta)$ and $\underline{VaR}_\beta(\mathcal{Z},\delta)$ be based on the data $\mathcal{Z}$. With a probability of at least $1-\delta$, it holds that
	\begin{align*}
	\underline{VaR}_\beta(\mathcal{Z},\delta)\le VaR_\beta(Z)\le \overline{VaR}_\beta(\mathcal{Z},\delta).
	\end{align*}
\end{proposition}

We remark that Theorem \ref{thm:mmm_} assumes that $F_Z$ is continuous. If $F_Z$ is not continuous, one can derive upper and lower bounds by using order statistics following  \citet[Lemma 3]{nikolakakis2021quantile}.

 \subsection{Conditional Value-at-Risk (CVaR)}
 \label{sec:cvar}
 For a risk level of $\beta\in(0,1)$, recall that the CVaR of $Z$ is given by
\begin{align*}
	    CVaR_\beta(Z):= \inf_{\alpha \in \mathbb{R}} \big(\alpha+(1-\beta)^{-1}E([Z-\alpha]^+)\big)
\end{align*} 
	where $[Z-\alpha]^+:=\max(Z-\alpha,0)$. For estimating  $CVaR_\beta(Z)$ from data $\mathcal{Z}$, we focus here on the case where the random variable $\rho^\phi(X,t)$ (and hence $Z$) has bounded support for fixed $t$. In particular, we assume that $P(\rho^\phi(X,t)\in[a,b])=1$. Note that $\rho^\phi(X,t)$ has bounded support when the function $\rho^\phi$ is bounded, which can be achieved  either by construction of $\phi$ or by clipping off $\rho^\phi$ outside the interval $[a,b]$ for some a priori chosen  constants $a$ and $b$, i.e., values outside this interval are clipped to the end points $a$ and $b$ of the interval. We remark that clipping off $\rho^\phi$ is not restrictive in most practical applications, i.e., realizations of $\rho^\phi(X,t)$ that are larger than a sufficiently large value of $b>0$ indicate robust satisfaction of $\phi$ and will not affect the risk associated with $Z$  while realizations of $\rho^\phi(X,t)$ smaller than $a<0$ violate the specification $\phi$ already.\footnote{In practice, it hence makes sense to select a negative value for $a$ and to select $b$ based on physical intuition that we may have - either from trajectories that we may have already observed or from domain knowledge, e.g., for a lane keeping controller  in autonomous driving the value of $b=1$ meter is a good robustness.} We will provide illustrative examples in our simulations in Section \ref{sec:simulations}. This boundedness assumption enables us now to directly leverage results from \citet{wang2010deviation} to estimate upper and lower bounds of $CVaR_\beta(Z)$. Let us first define the empirical estimate of $CVaR_\beta(Z)$ as
	\begin{align*}
	    \widehat{CVaR}_\beta(\mathcal{Z}):=\inf_{\alpha\in{\mathbb{R}}}\Big(\alpha+(N(1-\beta))^{-1}\sum_{i=1}^N[Z^i-\alpha]^+\Big).
	\end{align*}
	Based on \citet[Theorem 3.1]{wang2010deviation}, we can now calculate an upper bound of $CVaR_\beta(Z)$ as
	\begin{align*}
	    &\overline{CVaR}_\beta(\mathcal{Z},\delta):=\widehat{CVaR}_\beta(\mathcal{Z})+\sqrt{\frac{5\ln(3/\delta)}{N(1-\beta)}}(b-a)
	\end{align*}
	and a lower bound as
	\begin{align*}
	    &\underline{CVaR}_\beta(\mathcal{Z},\delta):=\widehat{CVaR}_\beta(\mathcal{Z})-\sqrt{\frac{11\ln(3/\delta)}{N(1-\beta)}}(b-a).
	\end{align*}
	We would like to highlight that the upper and lower bounds $\overline{CVaR}_\beta(\mathcal{Z},\delta)$ and $\underline{CVaR}_\beta(\mathcal{Z},\delta)$, respectively, become less accurate with larger values of $(b-a)$ which we can account for by increasing the number of observed trajectories $N$. The following proposition follows immediately from \citet[Theorem 3.1]{wang2010deviation}.
	\begin{proposition}\label{prop:2}
	    Let $\delta\in(0,1)$ be a probability threshold and $\beta\in(0,1)$ be a risk level. Assume that $P(\rho^\phi(X,t)\in[a,b])=1$. Let $\overline{CVaR}_\beta(\mathcal{Z},\delta)$ and $\underline{CVaR}_\beta(\mathcal{Z},\delta)$ be based on the data $\mathcal{Z}$. With a probability of at least $1-\delta$, it holds that
	\begin{align*}
	\underline{CVaR}_\beta(\mathcal{Z},\delta)\le CVaR_\beta(Z)\le \overline{CVaR}_\beta(\mathcal{Z},\delta).
	\end{align*}
	\end{proposition}
	
	\begin{remark}
	    The case where $Z$ has unbounded support, but where $Z$ is  sub-Gaussian or sub-exponential, has been considered in \citet{brown2007large,thomas2019concentration,kolla2019concentration,bhat2019concentration,mhammedi2020pac}. 
	\end{remark}

\subsection{Mean}
Define the empirical estimate of the mean $E(Z)$ as
\begin{align*}
\widehat{E}(\mathcal{Z}):=\frac{1}{N}\sum_{i=1}^NZ^i.
\end{align*}
By the law of large numbers,  $\widehat{E}(\mathcal{Z})$ converges to $E(Z)$ with probability one as $N$ goes to infinity. For finite $N$ and when again $Z$ has bounded support, i.e., $P(Z\in[a,b])=1$, we can apply Hoeffding's inequality and calculate an upper $\overline{E}(\mathcal{Z},\delta)$ of the mean $E(Z)$ as
\begin{align*}
    \overline{E}(\mathcal{Z},\delta)&:=\widehat{E}(\mathcal{Z})+\sqrt{\frac{\ln(2/\delta)}{2N}}(b-a)
\end{align*}
and a lower bound as
\begin{align*}
    \underline{E}(\mathcal{Z},\delta)&:=\widehat{E}(\mathcal{Z})-\sqrt{\frac{\ln(2/\delta)}{2N}}(b-a).
\end{align*}
Similarly to the observation that we made for CVaR, note that the upper and lower bounds $\overline{E}(\mathcal{Z},\delta)$ and $\underline{E}(\mathcal{Z},\delta)$, respectively, become less accurate with increasing values of $(b-a)$, and more accurate with increasing $N$. We next show that we indeed obtain valid upper and lower bounds.

\begin{proposition}\label{thm:mmmmm}
    Let $\delta\in(0,1)$ be a probability threshold. Assume that $P(\rho^\phi(X,t)\in[a,b])=1$. Let $\overline{E}(\mathcal{Z},\delta)$ and $\underline{E}(\mathcal{Z},\delta)$ be based on the data $\mathcal{Z}$. With a probability of at least $1-\delta$, it holds that
	\begin{align*}
	\underline{E}(\mathcal{Z},\delta)\le E(Z)\le \overline{E}(\mathcal{Z},\delta).
	\end{align*}
\end{proposition}

\begin{myexpcont2}
 \begin{figure*}
	\centering
	\includegraphics[scale=0.4125]{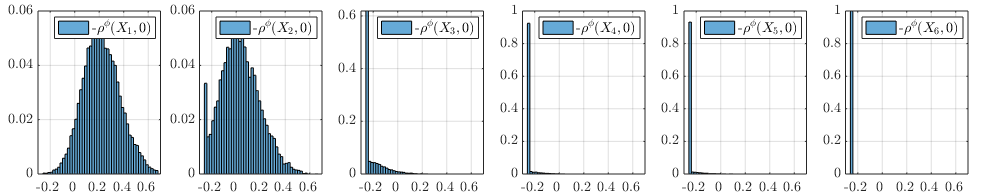}
	\caption{Histogram of  $-\text{RD}^{\phi}(X_j,0)$ of the specification $\phi$ in \eqref{ex:1_formula} for robot trajectories $j\in\{1,\hdots,6\}$.}
	\label{fig:ex11}
\end{figure*}
We now modify Example \ref{ex1} by considering that the regions $C$ and $D$  are not exactly known. Let  $c$ and $d$ in \eqref{eq:c_} and \eqref{eq:d_}, respectively, be Gaussian random vectors as
\begin{align}
    c\sim\mathcal{N}\Big(\begin{bmatrix}2 \\ 3 \end{bmatrix},\begin{bmatrix}0.2 & 0\\ 0 & 0.2\end{bmatrix}\Big),\label{distr_c}\\
    d\sim \mathcal{N}\Big(\begin{bmatrix}6 \\ 4 \end{bmatrix},\begin{bmatrix}0.2 & 0\\ 0 & 0.2\end{bmatrix}\Big)\label{distr_d}.
\end{align}
Consequently, the signals $x_1$-$x_6$ become stochastic processes denoted by $X_1$-$X_6$. Let now $X_j^i$ denote the $i$th observed realization of $X_j$ where $j\in\{1,\hdots,6\}$.  Our first goal is to estimate $VaR_\beta(Z)$ to compare the risk between the six robot trajectories $r_1$-$r_6$. We set $\delta:=0.01$ and $N:=15000$.\footnote{We can select smaller $N$ at the cost of slightly more conservative estimates.} The histograms of $-\rho^\phi(X_j)$ for each trajectory are shown in Fig. \ref{fig:ex11}. For different risk levels $\beta$, the resulting upper and lower bounds for the value-at-risk are shown in the next table. 
\begin{center}
\begin{tabular}{|p{1.3cm}|p{1cm}|p{1.2cm}|p{1cm}|p{1.2cm}|p{1cm}|p{1.2cm}|p{1cm}|p{1.2cm}|}
\hline
\backslashbox{$j$}{$R$} & $\overline{VaR}_{0.9}$ &  $\overline{VaR}_{0.925}$ &
$\overline{VaR}_{0.95}$ & 
$\overline{VaR}_{0.975}$ &
$\underline{VaR}_{0.9}$ &
$\underline{VaR}_{0.925}$ &  $\underline{VaR}_{0.95}$ &  $\underline{VaR}_{0.975}$ \\
\hline
1 & 0.434 & 0.467 & 0.508 & 0.577 & 0.407 & 0.432 & 0.465 & 0.505 \\
2 & 0.261 & 0.295 & 0.335 & 0.424 & 0.232 & 0.259 & 0.292 & 0.332 \\
3 & -0.075 & -0.044 & 0.001 & 0.086 & -0.1 & -0.077 & -0.046 & -0.003 \\
4 & -0.25 & -0.222 & -0.177 & -0.086 & -0.25 & -0.25 & -0.225 & -0.182 \\
5 & -0.249 & -0.228 & -0.18 & -0.084 & -0.249 & -0.249 & -0.23 & -0.185 \\
6 & -0.249 & -0.249 & -0.249 & -0.249 & -0.249 & -0.249 & -0.249 & -0.249 \\
\hline
\end{tabular}
\end{center}
Across all $\beta$, it can be observed that the estimate $\overline{VaR}_\beta$ of $VaR_\beta$ is relatively tight as the difference $|\overline{VaR}_\beta-\underline{VaR}_\beta|$ between upper  and lower bounds is small. The table indicates that trajectories $r_1$ and $r_2$ are not favorable and are not robust. Recall that smaller risk values are favorable as only negative values indicate actual robustness. Trajectory $r_3$ is better compared to trajectories $r_1$ and $r_2$, but worse than $r_4$-$r_6$ in terms of the approximate STL robustness risk of $\phi$. For trajectories $r_4$-$r_6$, note that a $\beta=0.9$ provides the information that the trajectories have roughly the same approximate STL robustness risk. However, once the risk level $\beta$ is increased to $0.925$, $0.95$, and $0.975$, it becomes clear that $r_6$ is preferable over $r_4$ and  $r_5$. This matches with what one would expect by closer inspection of Fig. \ref{ex:1_figure} and Fig. \ref{fig:ex11}.

We next estimate $CVaR_\beta(Z)$ and therefore restrict $\rho^\phi$ to lie within $[-0.5,0.25]$ simply by clipping values that exceed this bound. This choice is motivated by our previous discussion in Section \ref{sec:cvar} and as $\rho^\phi$ is upper bounded by $0.25$, see histograms in Fig. \ref{fig:ex11}. For different risk levels $\beta$, the resulting upper and lower bounds for the conditional value-at-risk are shown next.
\begin{center}
\begin{tabular}{|p{1.3cm}|p{1cm}|p{1.22cm}|p{1.1cm}|p{1.21cm}|p{1cm}|p{1.22cm}|p{1.1cm}|p{1.22cm}|}
\hline
\backslashbox{$j$}{$R$} & $\overline{CVaR}_{0.9}$ &  $\overline{CVaR}_{0.925}$ &
$\overline{CVaR}_{0.95}$ & 
$\overline{CVaR}_{0.975}$ &
$\underline{CVaR}_{0.9}$ &
$\underline{CVaR}_{0.925}$ &  $\underline{CVaR}_{0.95}$ &  $\underline{CVaR}_{0.975}$ \\
\hline
1 & 0.577 & 0.607 & 0.645 & 0.707 & 0.32 & 0.31 & 0.282 & 0.193 \\
2 & 0.432 & 0.471 & 0.527 & 0.637 & 0.175 & 0.174 & 0.164 & 0.12 \\
3 & 0.1 & 0.136 & 0.193 & 0.301 & -0.16 & -0.161 & -0.17 & -0.213 \\
4 & -0.078 & -0.04 & 0.019 & 0.13 & -0.335 & -0.336 & -0.344 & -0.384 \\
5 & -0.08 & -0.042 & 0.019 & 0.134 & -0.337 & -0.338 & -0.344 & -0.38 \\
6 & -0.146 & -0.13 & -0.103 & -0.042 & -0.403 & -0.426 & -0.466 & -0.556\\
\hline
\end{tabular}
\end{center}
In general, the same observations regarding the ranking of $r_1-r_6$ can be made based on the conditional value-at-risk. However, the risk levels are in general much higher as ${CVaR}_\beta$ is more risk sensitive than ${VaR}_\beta$. An important observation is that the estimates $\overline{CVaR}_\beta$ of $CVaR_\beta$ are not as tight as before for ${VaR}_\beta$ as the difference $|\overline{CVaR}_\beta-\underline{CVaR}_\beta|$ is larger, particularly for larger $\beta$ due to the division by $1-\beta$ in the estimates of $\overline{CVaR}_\beta$ and $\underline{CVaR}_\beta$. For completeness, we also report the estimated mean of $Z$.
\begin{center}
\begin{tabular}{|p{1.3cm}|p{1.2cm}|p{1.2cm}|}
\hline
\backslashbox{$j$}{$R$} & $\overline{E}$ &  $\underline{E}$ \\
\hline
1 & 0.227 & 0.207  \\
2 & 0.043  & 0.023 \\
3 & -0.194 & -0.214 \\
4 & -0.233 & -0.253 \\
5 & -0.233 & -0.253 \\
6 & -0.24 & -0.26 \\
\hline
\end{tabular}
\end{center}
\end{myexpcont2}

\section{Exact Computation of the Approximate STL Robustness Risk}
\label{sec:calc_risk}

In the previous section, we estimated the approximate STL robustness risk using observed realizations $X^1,\hdots,X^N$ of the stochastic process $X$. In this section, we instead assume to know the distribution of $X$. There are  two main challenges in computing the approximate STL robustness risk $R(-\rho^\phi(X,t))$ from the distribution of $X$.  First,  note that exact computation of $R(-\rho^\phi(X,t))$ requires knowledge of the CDF of $\rho^\phi(X,t)$.  However, the CDF of $\rho^\phi(X,t)$ is in general not known and often hard to obtain analytically. Second, calculating $R(-\rho^\phi(X,t))$ may often involve solving high dimensional integrals for which in most of the cases no closed-form expressions exists. For these reasons, we assume in this section that the STL formula $\phi$ is bounded and that $X:T\times \Omega \to \mathcal{X}$ is a discrete-time stochastic process, i.e., $T:=\mathbb{Z}$, with a finite state space $\mathcal{X}\subseteq \mathbb{R}^n$ (i.e., the set $\mathcal{X}$ consists of a finite set of elements).

Recall that the time intervals $I$ contained in a bounded STL formula $\phi$ are compact. The satisfaction of such an STL formula can hence be decided by finite signals. A bounded STL formula $\phi$ has a  future formula length $L^\phi_f\in\mathbb{Z}$ and a past formula length $L^\phi_p\in\mathbb{Z}$. The future formula length $L^\phi_f$ can be calculated, similarly to \citet{sadraddini2015robust}, as
 \begin{align*}
     L^\top_f&=L^\mu_f:=0\\
     L^{\neg\phi}_f&:=L^\phi_f\\
     L^{\phi'\wedge\phi''}_f&:=\max(L^{\phi'}_f,L^{\phi''}_f)\\
     L^{\phi' U_I \phi''}_f&:=\max \{I\cap \mathbb{Z}\}+\max(L^{\phi'}_f,L^{\phi''}_f)\\
     L^{\phi' \underline{U}_I \phi''}_f&:=\max(L^{\phi'}_f,L^{\phi''}_f).
 \end{align*}
 The past formula length $L^\phi_p$ can be calculated similarly as
  \begin{align*}
     L^\top_p&=L^\mu_p:=0\\
     L^{\neg\phi}_p&:=L^\phi_p\\
     L^{\phi'\wedge\phi''}_p&:=\max(L^{\phi'}_p,L^{\phi''}_p)\\
     L^{\phi' U_I \phi''}_p&:=\max(L^{\phi'}_p,L^{\phi''}_p)\\
     L^{\phi' \underline{U}_I \phi''}_p&:=\max \{I\cap \mathbb{Z}\}+\max(L^{\phi'}_p,L^{\phi''}_p).
 \end{align*}
 A finite signal of length $L^\phi_f+L^\phi_p$ is now sufficient to determine if $\phi$ is satisfied at time $t$. In particular, information from the time interval $T_L:=\{t-L^\phi_p,\hdots,t,\hdots,t+L^\phi_f\}$ is sufficient to determine if $\phi$ is satisfied at time $t$. Now, let $X:\Omega\times T_L\to \mathcal{X}$ be the discrete-time stochastic process under consideration where the state space $\mathcal{X}\subseteq \mathbb{R}^n$ is a finite set. Note that we can always obtain such a finite set $\mathcal{X}$ from a continuous state space by discretization. Let the probability mass function (PMF) $f_X(x)$ of $X$ be given. The next result is stated without proof as it follows immediately from the fact that $T_L$ and $\mathcal{X}$, and consequently the set of signals $\mathfrak{F}(T_L,\mathcal{X})$, are finite sets.

\begin{proposition}\label{prop_discr}
	Let $\phi$ be a bounded STL formula with future and past formula lengths $L^\phi_f$ and $L^\phi_p$, respectively. Let  $X:\Omega\times T_L\to \mathcal{X}$ be a discrete-time stochastic process with a finite state space $\mathcal{X}$.  For $t\in \mathbb{R}$, we can calculate the PMF $f_Z(z)$ and the CDF $F_Z(z)$   of $Z$ as
	\begin{align*}
	f_Z(z)=\sum_{x \in \mathfrak{F}(T_L,\mathcal{X})} \mathbb{I}(-\rho^\phi(x,t)= z)f_X(x),\\
	F_{Z}(z)=\sum_{x \in \mathfrak{F}(T_L,\mathcal{X})} \mathbb{I}(-\rho^\phi(x,t)\le z)f_X(x).\\
	\end{align*}
\end{proposition}
Note that $F_Z(z)=\sum_{z'\le z}f_Z(z')$ holds as required. Having obtained the PMF $f_Z(z)$ and the CDF $F_Z(z)$ of $Z$, it is now straightforward to calculate $R(Z)$ for various risk measures $R$. Note in particular that $Z$ is a discrete random variable so that $f_Z(z)$ is discrete and $F_Z(z)$ is piecewise continuous, hence simplifying the calculation of $R(Z)$ as no high-dimensional integrals need to be solved.

\begin{myexpcont}
Recall that $c$ and $d$ were assumed to be Gaussian distributed according to \eqref{distr_c} and \eqref{distr_d}, respectively. We first discretize the distributions of $c$ and $d$, see Appendix \ref{app:discretization} for details.  From the PMFs $f_c$ and $f_d$, we can now calculate the PMF $f_X(x)$ for any $x \in \mathfrak{F}(T_L,\mathbb{R}^6)\times\mathcal{C}\times\mathcal{D}$ where $\mathcal{C}$ and $\mathcal{D}$ are the discretized domains of $c$ and $d$. We can hence calculate $f_Z(z)$ according to Proposition \ref{prop_discr}. From this, the value at risk $VaR_\beta(Z)$ can be calculated which is reported in the next table.
 \begin{center}
    \begin{tabular}{|p{1.2cm}|p{1.2cm}|p{1.2cm}|p{1.2cm}|p{1.2cm}|}
    \hline
    \backslashbox{$i$}{$R$} & $VaR_{0.9}$ & $VaR_{0.925}$ & $VaR_{0.95}$ & $VaR_{0.975}$ \\
    \hline
    1 & 0.403 & 0.429 & 0.461 & 0.509 \\
    2 & 0.225 & 0.255 & 0.29 & 0.348 \\
    3 & -0.102 & -0.067 & -0.049 & 0.003 \\
    4 & -0.249 & -0.249 & -0.222 & -0.162\\
    5 & -0.25 & -0.25 & -0.222 & -0.157 \\
    6 & -0.249 & -0.249 & -0.249 & -0.249 \\
    \hline
    \end{tabular}
\end{center}
   It can be seen that the STL robustness risks reported above closely resemble the sampling-based estimates $\overline{VaR}_\beta$ of ${VaR}_\beta$ from Section \ref{comppp}.
\end{myexpcont}

%% file: chapter/simulations.tex
\section{Simulations: Autonomous Driving in Carla}
\label{sec:simulations}

We consider the verification of neural network-based lane keeping controllers for lateral control in the autonomous driving simulator CARLA \citep{dosovitskiy2017carla}, see  Fig. \ref{fig:CARLA_} (left). Lane keeping in CARLA is achieved by tracking a set of predefined waypoints.  For longitudinal control, a built-in PID controller is used to stabilize the car at 20 km/h. We particularly trained four different neural network controllers as detailed below.  Our overall goal is to estimate and compare the risks of these four controllers for five different specifications during a double left turn, see Fig. \ref{fig:CARLA_} (middle).  

For the verification and comparison of these controllers, we are particularly interested in the cross-track error, which is a measure of the closest distance from the car to the path defined by the set of waypoints as illustrated in Fig. \ref{fig:CARLA_} (right). Formally, let $wp_1$ be the waypoint that is closest to the car and let $wp_2$ be the waypoint proceeding $wp_1$. Then the cross-track error is defined as $c_e:=\|w\|\sin(\theta_w)$ where $w$ is the vector pointing from $wp_1$ to the car and $\theta_w$ is the angle between $w$ and the vector pointing from $wp_1$ to $wp_2$. We are also interested in the orientation error $\theta_e:=\theta_t-\theta$ between the orientation of the reference path $\theta_t$ and the orientation of the car $\theta$. 

The state $x:=(c_e,\theta_e,v,d,\dot{\theta}_t)$ of the car consists of the cross-track error $c_e$, the orientation error $\theta_e$, the velocity $v$ of the car, the internal state $d$ of the longitudinal PID controller, and the rate $\dot{\theta}_t$ at which the orientation of the reference path changes. The control input for which we aim to learn and verify a lane keeping controller is the steering angle $u$.

\subsection{Training Neural Network Lane Keeping Controllers}
We have trained four different neural network controllers. Two of these four controllers were obtained by using supervised imitation learning (IL) \citep{ross2010efficient}, while the other two controllers were obtained by learning control barrier functions (CBFs) from expert demonstrations \citep{lindemann2021learning}. 

To obtain  two imitation learning controllers, we used a CARLA built-in PID controller $u^*$ as an expert controller  to collect expert trajectories, which are sequences of state and control input pairs. The first IL controller, denoted as IL$_\text{full}$, is trained using the full state  $x$ as an input to the neural network, while the control input $u$ is the output. The second IL controller, denoted as IL$_\text{partial}$, is trained by only using partial state knowledge. In particular, only the cross-track error $c_e$, the orientation error $\theta_e$, and the rate $\dot{\theta}_t$ at which the orientation of the path changes are used here as an input to the neural network. We used one-layer neural networks with 20 neurons per layer and ReLU activation functions, and trained with the mean squared error as the loss function. 
\begin{remark}
For simplicity, we did not attempt to address the distribution shift between the expert controller and the trained controller, e.g., by using DAGGER \citep{ross2011reduction}. We remark that our primary goal lies in the verification and comparison of risk between controllers.
\end{remark}

To obtain the CBF-based controllers, we again used the expert controller $u^*$ to get expert trajectories from which we learned robust control barrier functions following \citet{lindemann2021learning}. The first controller, denoted as CBF$_\text{full}$, uses again full state knowledge of $x$. The second controller, denoted as CBF$_\text{partial}$, estimates the cross-track error $c_e$ from RGB dashboard camera images while assuming knowledge of the remaining states, see \citet{lindemann2021learning} for details. Both neural network controllers consist of two layers with 32 and 16 neurons and tanh activation functions.

\subsection{Risk Verification and Comparison}

\begin{figure*}
\centering
\includegraphics[scale=0.165]{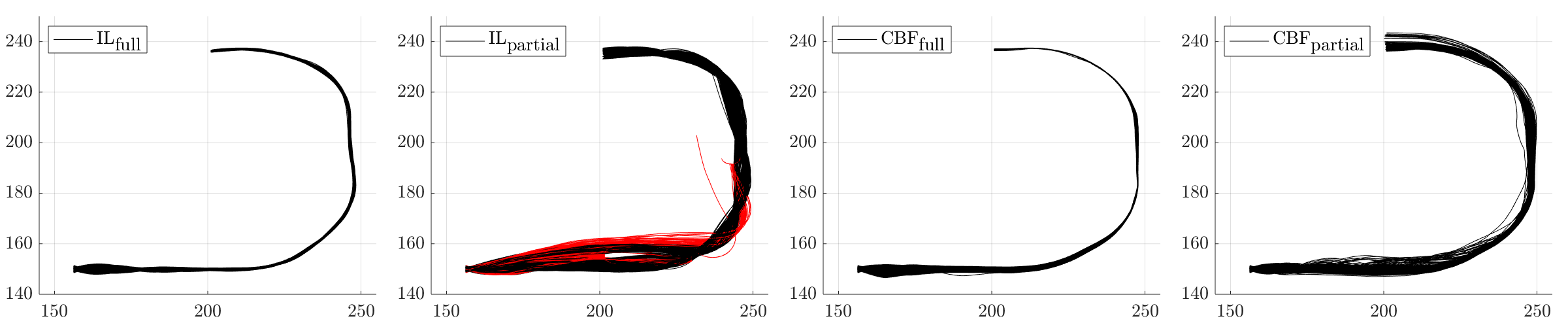}
\caption{Shown are 600 trajectories for each of the four controllers during the double left turn. Trajectories marked in red led to a collision with an obstacle.}
\label{fig:trajectories}
\end{figure*}

For the risk verification and comparison of these four controllers, we tested each of them on the training course, see Fig. \ref{fig:CARLA_} (middle). We uniformly sampled the initial position of the car in a range of $c_e\in [-1,1]$ m and $\theta_e\in [-0.4,0.4]$ rad and added normally distributed noise in a range of $[-0.1,0.1]$ rad to the control input to simulate actuation noise so that the car becomes a stochastic process $X$. We collected $N:=1000$ trajectories for each controller of which 600 are shown in Fig. \ref{fig:trajectories}. From a visual inspection, we can already see that the controllers that use full state knowledge (IL$_\text{full}$, CBF$_\text{full}$) outperform the controllers that only use partial state knowledge (IL$_\text{partial}$, CBF$_\text{partial}$). Videos of each controller from five different initial conditions are provided under \href{https://www.dropbox.com/sh/9hzrujc94jm2jbh/AAAxcIwrzIZyZfk_tSjtuB7oa?dl=0}{https://tinyurl.com/48xjf545}.

To obtain a more formal assessment, we next estimate the risk of each controller with respect to: 1) the cross-track error over the whole trajectory, during steady state, and during the transient phase,  2) the responsiveness of the controller, and 3) the orientation error. 

\subsubsection{Cross-track error} The specification that we look at here is that the cross-track error $c_e$ should always be within the interval $[-2.25,2.25]$, where $2.25$ is a threshold that we selected based on the cross-track error induced by the expert controller $u^*$. In STL language, we have
\begin{align*}
    \phi_1:=G_{[0,\infty)} (|c_e|\le 2.25).
\end{align*}

 \begin{figure*}
	\centering
	\begin{subfigure}[b]{0.94\textwidth}
    	\includegraphics[scale=0.2]{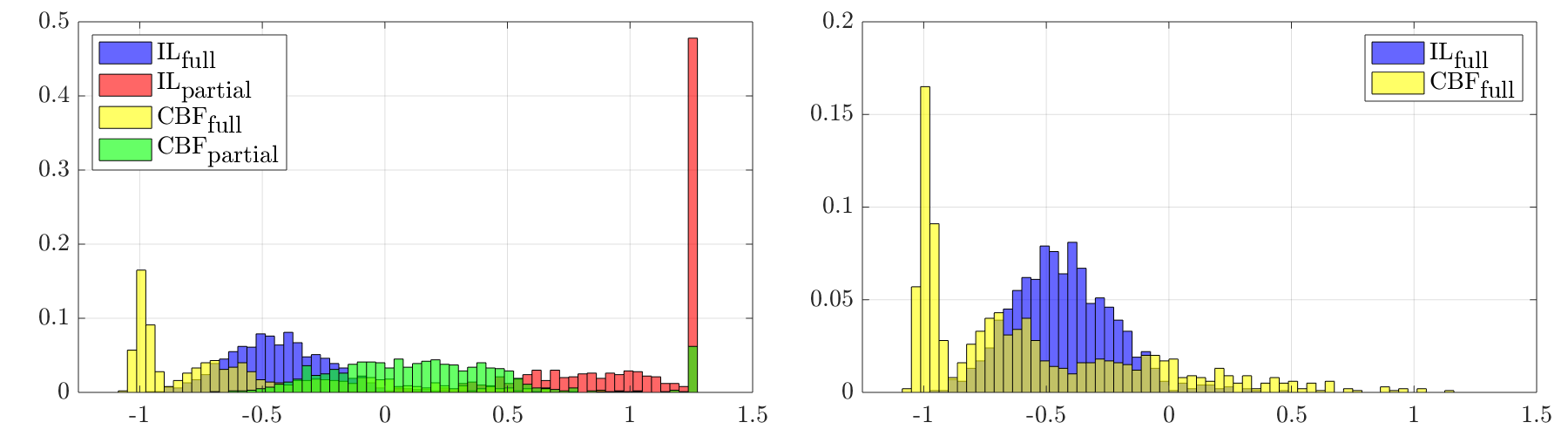}
    	\vspace{-0.2cm}
    	\caption{$\phi_1:=G_{[0,\infty)} (|c_e|\le 2.25)$}
        \label{fig:sim_1}
    \end{subfigure}
    \begin{subfigure}[b]{0.94\textwidth}
	    \includegraphics[scale=0.2]{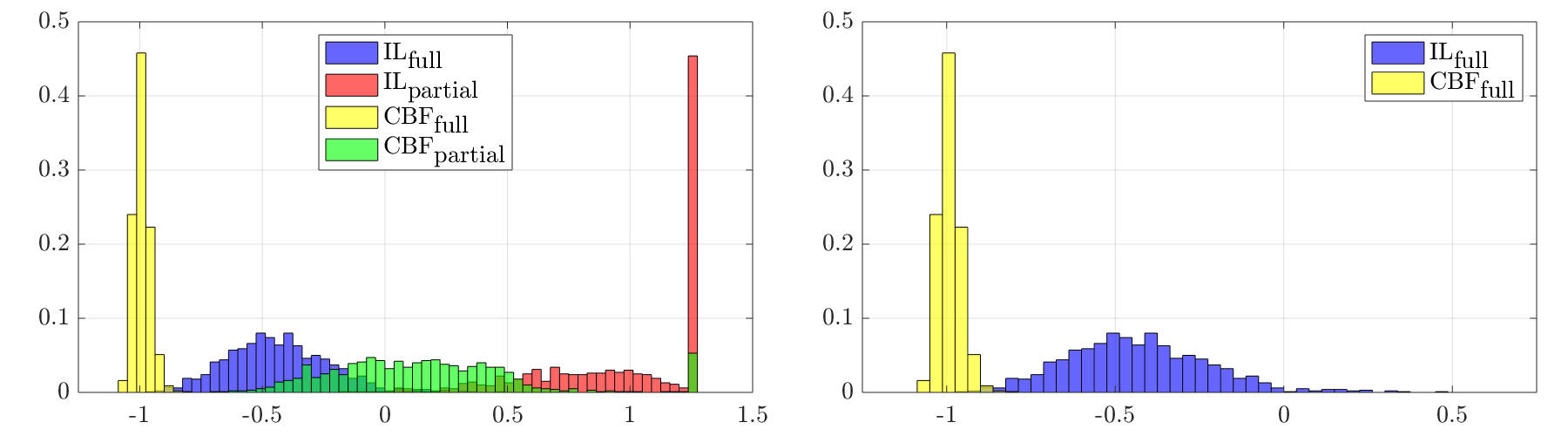}
	    \vspace{-0.2cm}
	    \caption{$\phi_2:=G_{[10,\infty)} (|c_e|\le 2.25)$}
        \label{fig:sim_2}
    \end{subfigure}
	\begin{subfigure}[b]{0.94\textwidth}
	    \includegraphics[scale=0.2]{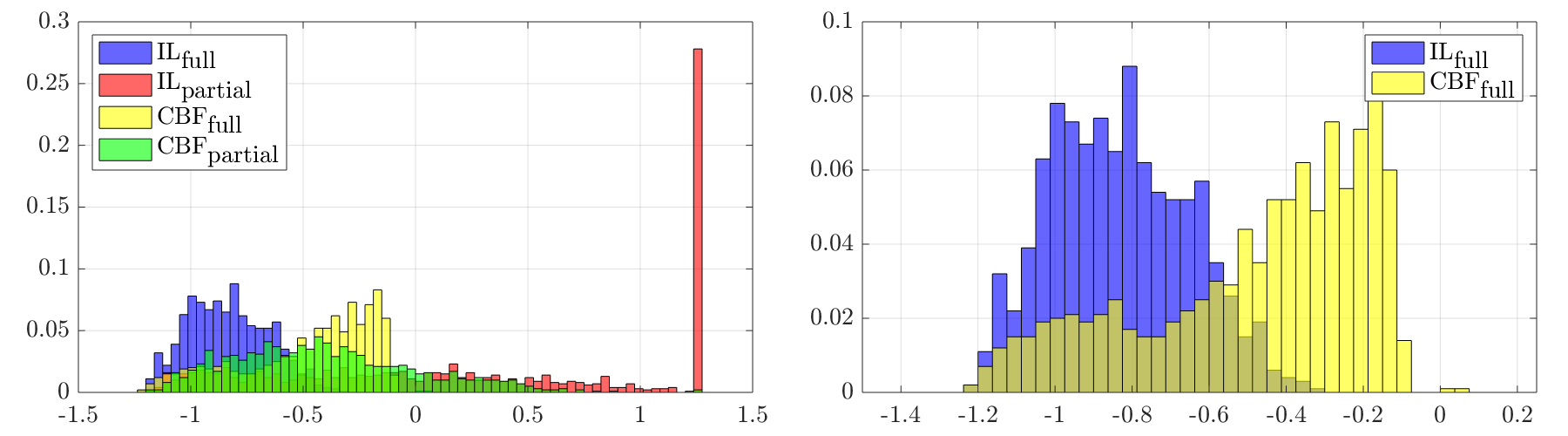}
	    \vspace{-0.2cm}
	    \caption{$\phi_3:=F_{[0,5]}G_{[0,5]}(|c_e|\le 1.25)$}
        \label{fig:sim_3}
    \end{subfigure}
	\begin{subfigure}[b]{0.94\textwidth}
	    \includegraphics[scale=0.2]{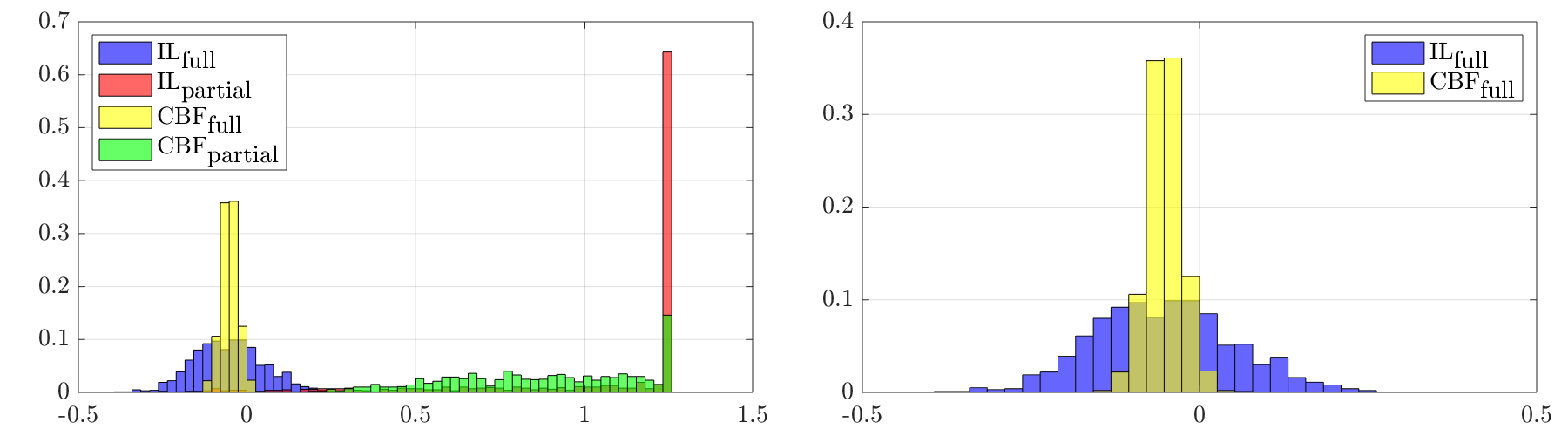}
	    \vspace{-0.2cm}
	    \caption{$\phi_4:=G_{[10,\infty)} \big((|c_e|\ge 1.25) \implies F_{[0,5]}G_{[0,5]}(|c_e|\le 1.25)\big)$}
        \label{fig:sim_4}
    \end{subfigure}
	\begin{subfigure}[b]{0.94\textwidth}
	    \includegraphics[scale=0.2]{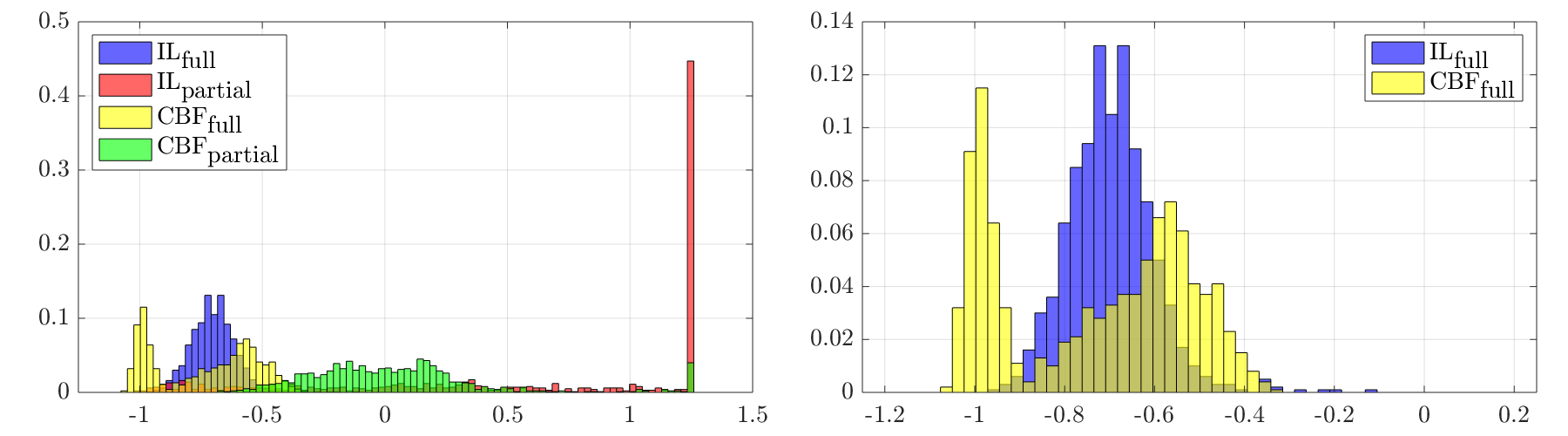}
	    \vspace{-0.2cm}
	    \caption{$\phi_5:=G_{[0,\infty)} \big((c_e\ge 1.25) \implies F_{[0,2]}G_{[0,1]}(\theta_e\le 0)\wedge (c_e\le -1.25) \implies F_{[0,2]}G_{[0,1]}(\theta_e\ge 0)\big)$}
        \label{fig:sim_5}
    \end{subfigure}
    \vspace{-0.15cm}
    \caption{Histograms of $-\rho^{\phi_i}(X,0)$ for each controller for the specifications $\phi_1$-$\phi_5$.}
\end{figure*}

We show the histograms of $\rho^{\phi_1}(X,0)$ for each controller in Fig. \ref{fig:sim_1} (left).\footnote{We restrict $\rho^{\phi_1}$ to lie within the interval $[-1.25,2.25]$, i.e., in this case we clip the values of $\rho^{\phi_1}(X,0)=\inf_{t\in\mathbb{Z}} 2.25-|c_e(t)|$ to $-1.25$  if $\rho^{\phi_1}(X,0)<-1.25$. In the remainder, we clip $\rho^{\phi_2}$-$\rho^{\phi_5}$ in the same way  for the specifications $\phi_2$-$\phi_5$.} We are particularly interested in the controllers IL$_\text{full}$ and CBF$_\text{full}$ and show their  histograms isolated in Fig. \ref{fig:sim_1} (right) for better readability. Selecting $\delta:=0.01$, the estimates of ${VaR}_{0.85}$, ${VaR}_{0.95}$, ${CVaR}_{0.85}$, and $E$ are reported in the table below. In the last column, we have additionally reported the empirical probability that the specification $\phi_1$ is satisfied which we calculate as
\begin{align*}
    \#_{\phi_1}:=\frac{\sum_{i=1}^N\mathbb{I}(\beta^{\phi_1}(X^i,0)=\top)}{N}.
\end{align*} For each risk measure,  we highlight the controller with the lowest risk in green. 
\begin{center}
\begin{tabular}{|p{1.3cm}|p{1cm}|p{1cm}|p{1.2cm}|p{0.8cm}|p{1cm}|p{1cm}|p{1.2cm}|p{0.8cm}|p{0.8cm}|}
\hline
\backslashbox{$u$}{$R$} & $\overline{VaR}_{0.85}$ &  $\overline{VaR}_{0.95}$ & $\overline{CVaR}_{0.85}$  &  $\overline{E}$  &
$\underline{VaR}_{0.85}$ & 
$\underline{VaR}_{0.95}$ &
$\underline{CVaR}_{0.85}$ &
$\underline{E}$ &
$\#_{\phi_1}$ \\
\hline
IL$_\text{full}$ & \textcolor{green}{-0.168} & \textcolor{green}{0.462} & \textcolor{green}{1.436} & -0.248 & -0.258 & -0.168 & -2.354 & -0.61 & \textcolor{green}{0.975}  \\
IL$_\text{partial}$ & 1.25 & 1.25 & 2.776 & 1.166 & 1.25 & 1.25 & -1.014 & 0.806 & 0.005 \\
CBF$_\text{full}$ & 0.135 & 1.125 & 1.818 & \textcolor{green}{-0.375} & -0.125 & 0.105 & -1.972 & -0.736 & 0.863 \\
CBF$_\text{partial}$ & 0.58 & 1.25 & 2.42 & 0.357 & 0.44 & 0.58 & -1.37 & -0.003 & 0.364 \\
\hline
\end{tabular}
\end{center}

Based on these risk estimates, we make the following observations:
\begin{itemize}
\item As expected from the visual inspection of Fig. \ref{fig:trajectories}, the  controllers IL$_\text{partial}$ and CBF$_\text{partial}$ perform poorly. Among these two, CBF$_\text{partial}$ performs slightly better in terms of risk than IL$_\text{partial}$. 
\item The controllers IL$_\text{full}$ and CBF$_\text{full}$ perform better. The risk of CBF$_\text{full}$ in terms of the expected value $\overline{E}$ is smaller than the risk of IL$_\text{full}$. Interestingly, the risk of IL$_\text{full}$ in terms of the $\overline{VaR}_{0.85}$,  $\overline{VaR}_{0.95}$, and $\overline{CVaR}_{0.85}$ is smaller than the risk of CBF$_\text{full}$. This is due to the long tail induced by CBF$_\text{full}$, see Fig. \ref{fig:sim_1} (right). We  hence argue that IL$_\text{full}$ is the better choice with respect to $\phi_1$.
\item The estimate $\overline{CVaR}_{0.85}$ of ${CVaR}_{0.85}$ is not tight and very conservative. The difference $|\overline{CVaR}_{0.85}-\underline{CVaR}_{0.85}|$ between the upper and lower bounds is large. To make this bound tighter, more data $N$ is needed. We neglect the conditional value-at-risk in the remainder.
\item In this case, it can be observed that a low empirical satisfaction probability $\#_{\phi_1}$ correlates with a high risk. We remark that this is not always the case as risk considers characteristics of the right tail of the distribution $-\rho^{\phi_1}(X,0)$, while satisfaction probabilities focus on the left tail of this distribution. This can be observed when we present the results for specification $\phi_5$. 
\end{itemize}

We formulate the hypothesis that the long tail of CBF$_\text{full}$ that makes CBF$_\text{full}$ more risky than IL$_\text{full}$ is induced by the transient behavior. We analyze this hypothesis in detail in the remainder looking at the specifications $\phi_2$ (steady-state) and $\phi_3$ (transient phase).

\subsubsection{Steady-state}

 In the  previous section, we concluded that IL$_\text{full}$ is the best controller for the specification $\phi_1$, i.e., when considering the cross-track error $c_e$ over the whole trajectory. We now study the steady-state behavior of each controller in terms of $c_e$ and reveal that  CBF$_\text{full}$ is the least risky controller when only looking at the steady-state. Therefore, we check if the cross-track error $c_e$ is always within the interval $[-2.25,2.25]$ after $10$ s by the specification
 \begin{align*}
     \phi_2:=G_{[10,\infty)} (|c_e|\le 2.25).
 \end{align*}
 We show the histograms of $\rho^{\phi_2}(X,0)$ for each controller Fig. \ref{fig:sim_2} and report the  risk estimates below. 
 \begin{center}
\begin{tabular}{|p{1.3cm}|p{1cm}|p{1cm}|p{1cm}|p{0.8cm}|p{0.8cm}|}
\hline
\backslashbox{$u$}{$R$} & $\overline{VaR}_{0.85}$ &  $\overline{VaR}_{0.9}$ & $\overline{VaR}_{0.95}$  &  $\overline{E}$ & $\#_{\phi_2}$  \\
\hline
IL$_\text{full}$ & -0.168 & -0.078 & 0.462 & -0.254 & 0.975  \\
IL$_\text{partial}$ & 1.25 & 1.25 & 1.25 & 1.153 & 0.005 \\
CBF$_\text{full}$ & \textcolor{green}{-0.944} & \textcolor{green}{-0.924} & \textcolor{green}{-0.794} & \textcolor{green}{-0.81} & \textcolor{green}{1} \\
CBF$_\text{partial}$ & 0.56 & 1.25 & 1.25 & 0.341 & 0.377 \\
\hline
\end{tabular}
\end{center}
 
 Based on these risk estimates, we make the following observations:
 \begin{itemize}
     \item  We see that our stated hypothesis is true and observe that CBF$_\text{full}$ now has the least risky behavior for all risk measures with respect to $\phi_2$, i.e., during steady state.
      \item For CBF$_\text{full}$, we have $\overline{VaR}_{0.95}(-\rho^{\phi_2}(X,0))=-0.794$. Consequently, for at most $5$ percent of the realizations the robustness is less than $0.794$.
 \end{itemize}

\subsubsection{Transient phase} Complementary to the previous analysis, we now look at the transient behavior of the cross-track error $c_e$ of each controller by imposing the specification
\begin{align*}
    \phi_3:=F_{[0,5]}G_{[0,5]}(|c_e|\le 1.25).
\end{align*}
In other words, the specification $\phi_3$ requires that eventually within the first $5$ s the absolute value of the cross-track error falls below the threshold $1.25$ for at least $5$ s. We show the histogram of each controller in Fig.  \ref{fig:sim_3} and report the corresponding risk estimates next.
 \begin{center}
\begin{tabular}{|p{1.3cm}|p{1cm}|p{1cm}|p{1cm}|p{0.8cm}|p{0.8cm}|}
\hline
\backslashbox{$u$}{$R$} & $\overline{VaR}_{0.85}$ &  $\overline{VaR}_{0.9}$ & $\overline{VaR}_{0.95}$  &  $\overline{E}$ & $\#_{\phi_3}$ \\
\hline
IL$_\text{full}$ & \textcolor{green}{-0.584} & \textcolor{green}{-0.524} & \textcolor{green}{-0.324} & \textcolor{green}{-0.652} & \textcolor{green}{1}  \\
IL$_\text{partial}$ & 1.25 & 1.25 & 1.25 & 0.493 & 0.42\\
CBF$_\text{full}$ & -0.157 & -0.137 & 0.063 & -0.297 & 0.998 \\
CBF$_\text{partial}$ & 0.2 & 0.38 & 1.25 & -0.221 & 0.83  \\
\hline
\end{tabular}
\end{center}

For $\phi_3$, we see a similar result as for $\phi_1$ in the sense that IL$_\text{full}$ is the least risky controller, but now clearly indicating that IL$_\text{full}$ is the less risky controller across all risk measures. It is also worth pointing out that CBF$_\text{full}$ and CBF$_\text{partial}$ have almost the same expected value, while $\overline{VaR}_{0.85}$, $\overline{VaR}_{0.9}$, and $\overline{VaR}_{0.95}$  indicate that CBF$_\text{full}$ is less risky.  

Summarizing the observations from $\phi_1$, $\phi_2$, and $\phi_3$,  IL$_\text{full}$ is the least risky controller during the transient phase and CBF$_\text{full}$ is the least risky controller during steady-state.

\subsubsection{Responsiveness} So far, we  focused on the cross-track error during steady-state and transient phase. We now analyze the responsiveness of the controllers when the cross-track error gets too large. We particularly analyze how responsive the controllers are in such situations and how quickly they can decrease the error again to an acceptable level. Let us therefore look at the specification
\begin{align*}
    \phi_4:=G_{[10,\infty)} \big((|c_e|\ge 1.25) \implies F_{[0,5]}G_{[0,5]}(|c_e|\le 1.25)\big).
\end{align*}
In other words, whenever the cross-track error $c_e$ leaves the interval $[-1.25,1.25]$ after the transient phase has died out (approximately after $10$ s), it should hold that within the next $5$ s the cross-track error is again within the interval $[-1.25,1.25]$ for at least $5$ s.  We show the histogram of each controller in Fig. \ref{fig:sim_4} and report the corresponding risk estimates below.
 \begin{center}
\begin{tabular}{|p{1.3cm}|p{1cm}|p{1cm}|p{1cm}|p{0.8cm}|p{0.8cm}|}
\hline
\backslashbox{$u$}{$R$} & $\overline{VaR}_{0.85}$ &  $\overline{VaR}_{0.9}$ & $\overline{VaR}_{0.95}$  &  $\overline{E}$ & $\#_{\phi_4}$ \\
\hline
IL$_\text{full}$ & 0.088 & 0.128 & 0.248 & \textcolor{green}{0.127} & 0.703  \\
IL$_\text{partial}$ & 1.25 & 1.25 & 1.25 & 1.226 & 0.026 \\
CBF$_\text{full}$ & \textcolor{green}{-0.0152} & \textcolor{green}{-0.005} & \textcolor{green}{0.055} & 0.129 & \textcolor{green}{0.974} \\
CBF$_\text{partial}$ & 1.25 & 1.25 & 1.25 & 1.054 & 0 \\
\hline
\end{tabular}
\end{center}

The results are interesting in the sense that the risk of IL$_\text{full}$ and CBF$_\text{full}$ in terms of the expected value are almost identical, even slightly favoring IL$_\text{full}$,  while the risk of CBF$_\text{full}$ in terms of $\overline{VaR}_{0.85}$, $\overline{VaR}_{0.9}$, and $\overline{VaR}_{0.95}$ is much smaller.

\subsubsection{Orientation Error}
Let us now focus on the orientation error $\theta_e$. In general, an orientation error is expected when either the  orientation $\theta_t$ of the reference path changes or the car tries to reduce the cross-track error $c_e$ by adjusting $\theta$, e.g., when $|c_e|>0$ we need $|\theta_e|>0$ to reduce $|c_e|$ (see Fig. \ref{fig:CARLA_}). To analyze how well the orientation error is adjusted when the cross-track error leaves the interval $[-1.25,1.25]$, we consider the specification
\begin{align*}
    \phi_5:=G_{[0,\infty)} \big(&(c_e\ge 1.25) \implies F_{[0,2]}G_{[0,1]}(\theta_e\le 0)\wedge (c_e\le -1.25) \implies F_{[0,2]}G_{[0,1]}(\theta_e\ge 0)\big).
\end{align*}
The specification $\phi_5$ encodes that, whenever the cross-track error $c_e$ leaves the interval $[-1.25,1.25]$, the orientation error $\theta_e$ should, within $2$ s, be such that the cross-track error decreases for at least $1$ s. We show the histogram of each controller in Fig. \ref{fig:sim_5} and report the   risk estimates below. 

 \begin{center}
\begin{tabular}{|p{1.3cm}|p{1cm}|p{1cm}|p{1cm}|p{0.8cm}|p{0.8cm}|}
\hline
\backslashbox{$u$}{$R$} & $\overline{VaR}_{0.85}$ &  $\overline{VaR}_{0.9}$ & $\overline{VaR}_{0.95}$  &  $\overline{E}$ & $\#_{\phi_5}$ \\
\hline
IL$_\text{full}$ & \textcolor{green}{-0.58} & \textcolor{green}{-0.54} & -0.13 & -0.517  & \textcolor{green}{1} \\
IL$_\text{partial}$ & 1.25 & 1.25 & 1.25 & 0.762 & 0.247 \\
CBF$_\text{full}$ & -0.47& -0.44 & \textcolor{green}{-0.32} & \textcolor{green}{-0.553} & \textcolor{green}{1}  \\
CBF$_\text{partial}$ & 0.43 & 1.14 & 1.25 & 0.225 & 0.503 \\
\hline
\end{tabular}
\end{center}

We can observe that the risk of IL$_\text{full}$ is the lowest for $\overline{VaR}_{0.85}$ and $\overline{VaR}_{0.9}$, while the risks of IL$_\text{full}$ and CBF$_\text{full}$ are roughly equal for the expected value $\overline{E}$. However, the distribution induced by IL$_\text{full}$ has a long tail which is why the risk of CBF$_\text{full}$ is the lowest for $\overline{VaR}_{0.95}$. 
%The identified car model is taken from  \citet{lindemann2021learning}. In particular, a local model in error coordinates with respect to the waypoints is given as
%\begin{align*}
%    \dot{v}&=-1.0954v-0.007v^2-0.1521d+3.7387\\
%    \dot{d}&=3.6v-20\\
%    \dot{c}_e &=v\sin(\theta_e),\\
%    \dot{\theta}_e&=v/2.51\tan(\delta)-\dot{\theta}_t
%\end{align*}
%where $v$ is the velocity of the car, $d$ is the integrator state of the longitudinal PID controller, $c_e$ is the cross-track error (defined below), $t{\theta}_t$ is the change in orientation of the reference path, and $\delta$ is the control input, i.e., the steering angle of the car. 

%% file: chapter/conclusion.tex
\section{Conclusion}
\label{sec:conclusion}

We defined the STL robustness risk to quantify the risk of a stochastic system lacking robustness against failure of an STL specification.  The approximate STL robustness risk was defined as a computationally tractable upper bound of the STL robustness risk. It was shown how the approximate STL robustness risk is estimated from data for the value-at-risk and the conditional value-at-risk. We also provided conditions under which the approximate STL robustness risk can be computed exactly. Within the autonomous driving simulator CARLA, we trained four different neural network lane keeping controllers and estimated their risk for five different STL system specifications.

%% file: chapter/appendix.tex
\section{Semantics of Signal Temporal Logic}
\label{app:STL}
The satisfaction function $\beta^\phi(x,t)$ determines whether or not the signal $x$ satisfies the specification $\phi$ at time $t$. The definition of $\beta^\phi(x,t)$ follows recursively from the structure of $\phi$ as follows.
\begin{definition}[STL Semantics]\label{def:qualitative_semantics}
For a signal $x:T\to\mathbb{R}^n$ and an STL formula $\phi$, the satisfaction function $\beta^\phi(x,t)$ is recursively defined as
	\begin{align*}
	\beta^\top(x,t)&:=\top,  \\
	\beta^\mu(x,t)&:=\begin{cases}
	\top &\text{ if }	x(t)\in O^\mu\\	
	\bot &\text{ otherwise, }	
	\end{cases}\\
	\beta^{\neg\phi}(x,t)&:= \neg \beta^{\phi}(x,t),\\
	\beta^{\phi' \wedge \phi''}(x,t)&:=\min(\beta^{\phi'}(x,t),\beta^{\phi''}(x,t)),\\
	%	(x,t) \models \phi' \vee \phi'' &\text{ iff } (x,t) \models \phi' \vee (x,t) \models \phi''\\
	\beta^{\phi' U_I \phi''}(x,t)&:=\sup_{t''\in (t\oplus I)\cap T}\Big( \min\big(\beta^{\phi''}(x,t''),\inf_{t'\in(t,t'')\cap T}\beta^{\phi'}(x,t')\big)\Big),\\
	\beta^{\phi' \underline{U}_I \phi''}(x,t)&:=\sup_{t''\in (t\ominus I)\cap T}\Big( \min\big(\beta^{\phi''}(x,t''),\inf_{t'\in(t'',t)\cap T}\beta^{\phi'}(x,t')\big)\Big).
	\end{align*}
\end{definition}

The semantics in Definition \ref{def:qualitative_semantics}, use the strict non-matching versions $U_I$ and $\underline{U}_I$ of the until operators. The non-strict matching versions of the until operator, in comparison, replace the open time intervals $(t,t'')$ in Definition \ref{def:qualitative_semantics} by the closed time intervals $[t,t'']$ as follows
\begin{align*}
	\beta^{\phi' \vec{U}_I \phi''}(x,t)&:=\sup_{t''\in (t\oplus I)\cap T}\Big( \min\big(\beta^{\phi''}(x,t''),\inf_{t'\in[t,t'']\cap T}\beta^{\phi'}(x,t')\big)\Big),\\
	\beta^{\phi' \vec{\underline{U}}_I \phi''}(x,t)&:=\sup_{t''\in (t\ominus I)\cap T}\Big( \min\big(\beta^{\phi''}(x,t''),\inf_{t'\in[t'',t]\cap T}\beta^{\phi'}(x,t')\big)\Big).
	\end{align*}

\section{Proof of Theorem \ref{thm:1}} 
We prove the statement of Theorem \ref{thm:1} first for the the semantics $\beta^\phi(X,t)$, then for the robust semantics  $\rho^\phi(X,t)$, and finally for the robustness degree $\text{RD}^{\phi}(X,t)$.

\subsection{Semantics $\beta^\phi(X,t)$} Let us define the power set of  $\mathbb{B}$ as 
$2^\mathbb{B}:=\{\emptyset,\top,\bot,\{\bot,\top\}\}$. Note that $2^\mathbb{B}$ is a  $\sigma$-algebra of $\mathbb{B}$. To prove measurability of $\beta^\phi(X(\cdot,\omega),t)$ in $\omega$  for a fixed $t\in T$, we need to show that, for each $B\in2^\mathbb{B}$, it holds that the inverse image  of $B$ under $\beta^\phi(X(\cdot,\omega),t)$ for a fixed $t\in T$ is contained within $\mathcal{F}$, i.e., that it holds that 
\begin{align*}
    \{\omega\in\Omega| \beta^\phi(X(\cdot,\omega),t)\in B\}\subseteq\mathcal{F}.
\end{align*} 
We  show measurability of $\beta^\phi(X(\cdot,\omega),t)$ in $\omega$ for a fixed $t\in T$ inductively on the structure of $\phi$. 

$\top$: For $B\in2^\mathbb{B}$, it trivially holds that $\{\omega\in\Omega| \beta^\top(X(\cdot,\omega),t)\in B\}\subseteq\mathcal{F}$ since $\beta^\top(X(\cdot,\omega),t)=\top$ for all $\omega\in\Omega$. This follows according to Definition \ref{def:qualitative_semantics} so that $\{\omega\in\Omega| \beta^\top(X(\cdot,\omega),t)\in B\}=\emptyset\subseteq\mathcal{F}$ if $B\in\{\emptyset,\bot\}$ and $\{\omega\in\Omega| \beta^\top(X(\cdot,\omega),t)\in B\}=\Omega\subseteq\mathcal{F}$ otherwise. 

$\mu$: Let  $1_{O^\mu}:\mathbb{R}^n\to\mathbb{B}$ be the indicator function of $O^\mu$ with $1_{O^\mu}(\zeta):=\top$ if  $\zeta\in O^\mu$ and $1_{O^\mu}(\zeta):=\bot$ otherwise. According to Definition \ref{def:qualitative_semantics}, we can now write $\beta^{\mu}(X(\cdot,\omega),t)=1_{O^\mu}(X(t,\omega))$.  Recall that $O^\mu$ is  measurable and note that the indicator function of a measurable set is measurable again (see e.g., \citet[Chapter 1.2]{durrett2019probability}). Since $X(t,\omega)$ is measurable in $\omega$ for a fixed $t\in T$ by definition, it  follows that $1_{O^\mu}(X(t,\omega))$ and hence $\beta^{\mu}(X(\cdot,\omega),t)$ is measurable in $\omega$ for a fixed $t\in T$.  In other words, for $B\in2^\mathbb{B}$, it  follows that 
\begin{align*}\{\omega\in\Omega| \beta^{\mu}(X(\cdot,\omega),t)\in B\}=\{\omega\in\Omega|1_{O^\mu}(X(t,\omega))\in B\}\subseteq \mathcal{F}.
\end{align*}

$\neg\phi$: By the induction assumption, $\beta^{\phi}(X(\cdot,\omega),t)$ is measurable in $\omega$  for a fixed $t\in T$. Recall that $\mathcal{F}$ is a $\sigma$-algebra that is, by definition, closed under its complement so that, for $B\in2^\mathbb{B}$, it holds that  
\begin{align*}\{\omega\in\Omega| \beta^{\neg\phi}(X(\cdot,\omega),t)\in B\}=\Omega\setminus \{\omega\in\Omega| \beta^{\phi}(X(\cdot,\omega),t)\in B\}\subseteq \mathcal{F}.
\end{align*}

$\phi'\wedge\phi''$: By the induction assumption, $\beta^{\phi'}(X(\cdot,\omega),t)$ and $\beta^{\phi''}(X(\cdot,\omega),t)$ are measurable in $\omega$  for a fixed $t\in T$. Hence $\beta^{\phi'\wedge\phi''}(X(\cdot,\omega),t)=\min(\beta^{\phi'}(X(\cdot,\omega),t),\beta^{\phi''}(X(\cdot,\omega),t))$ is measurable in $\omega$ for a fixed $t\in T$ since the min operator of measurable functions is again a measurable function.

$\phi' U_I \phi''$ and $\phi' \underline{U}_I \phi''$: Recall the definition of the future until operator 
\begin{align*}
    \beta^{\phi' U_I \phi''}(X(\cdot,\omega),t) := \underset{t''\in (t\oplus I)\cap T}{\text{sup}}  \big(\min(\beta^{\phi''}(X(\cdot,\omega),t''),\underset{t'\in (t,t'')\cap T}{\text{inf}}\beta^{\phi'}(X(\cdot,\omega),t') )\big).
\end{align*} 
By the induction assumption, $\beta^{\phi'}(X(\cdot,\omega),t)$ and $\beta^{\phi''}(X(\cdot,\omega),t)$ are measurable in $\omega$ for a fixed $t\in T$. First note that $(t,t'')\cap T$ and $(t\oplus I)\cap T$ are countable sets since $T=\mathbb{N}$. According to \citet[Theorem 4.27]{guide2006infinite}, the supremum and infimum operators over a countable number of measurable functions is again measurable.  Consequently, the function  $\beta^{\phi' U_I \phi''}(X(\cdot,\omega),t)$ is measurable in $\omega$ for a fixed $t\in T$. The same reasoning applies to $\beta^{\phi' \underline{U}_I \phi''}(X(\cdot,\omega),t)$. 

\subsection{Robust semantics $\rho^\phi(X,t)$} The proof for $\rho^\phi(X(\cdot,\omega),t)$ follows again inductively on the structure of $\phi$ and the goal is to show that  $\{\omega\in\Omega| \rho^\phi(X(\cdot,\omega),t)\in B\}\subseteq\mathcal{F}$ for each Borel set $B\in\mathcal{B}$. The difference here, compared to the proof for the semantics $\beta^\phi(X(\cdot,\omega),t)$ presented above, lies only in the way predicates $\mu$ are handled. Note first that we can write $\rho^\mu(X(\cdot,\omega),t)$ as
\begin{align}\label{eq:rho_mu}
\begin{split}
\rho^\mu(X(\cdot,\omega),t)&=0.5( 1_{O^\mu}(X(t,\omega))+1)\bar{d}(X(t,\omega),\text{cl}(O^{\neg\mu}))\\
&\hspace{2cm}+0.5(1_{O^\mu}(X(t,\omega))-1) \bar{d}(X(t,\omega),\text{cl}(O^\mu)).
\end{split}
\end{align} 
where we recall that we interpret $\top:=1$ and $\bot=-1$. Since the composition of the indicator function with $X(t,\omega)$, i.e.,  $1_{O^\mu}(X(t,\omega))$, is  measurable in $\omega$ for a fixed $t\in T$ as argued before, we only need to  show that $ \bar{d}(X(t,\omega),\text{cl}(O^\mu))$ and  $\bar{d}(X(t,\omega),\text{cl}(O^{\neg\mu}))$ are measurable in $\omega$ for a fixed $t\in T$. This immediately follows since $X(t,\omega)$ is measurable in $\omega$ for a fixed $t\in T$ by definition  and since the function $\bar{d}$ is continuous in its first argument, and hence  measurable (see \citet[Corollary 4.26]{guide2006infinite}), due to $d$ being a metric defined on the set $\mathbb{R}^n$ (see e.g., \citet[Chapter 3]{munkres1975prentice}) so that $\rho^\mu(X(\cdot,\omega),t)$ is measurable in $\omega$ for a fixed $t\in T$.

\subsection{Robustness Degree $\text{RD}^{\phi}(X,t)$}
For $\text{RD}^\phi(X(\cdot,\omega),t)$, note that, for a fixed $t\in T$, the function $\text{RD}^\phi$ maps from the domain $\mathfrak{F}(T,\mathbb{R}^n)$ into the domain $\mathbb{R}$, while $X(\cdot,\omega)$ maps from the domain $\Omega$ into the domain $\mathfrak{F}(T,\mathbb{R}^n)$. Recall now that $\text{RD}^\phi(X(\cdot,\omega),t)=\bar{\kappa}\big(X(\cdot,\omega),\text{cl}(\mathcal{L}^\phi(t))\big):=\inf_{x^*\in \text{cl}(\mathcal{L}^\phi(t))}\kappa(X(\cdot,\omega),x^*)$ and that $\kappa$ is a metric defined on the set $\mathfrak{F}(T,\mathbb{R}^n)$ as argued in \citet{fainekos2009robustness}. Therefore, it follows that the function $\bar{\kappa}$ is continuous in its first argument (see e.g., \citet[Chapter 3]{munkres1975prentice}), and hence measurable with respect to the Borel $\sigma$-algebra of $\mathfrak{F}(T,\mathbb{R}^n)$ (see e.g., \citet[Corollary 4.26]{guide2006infinite}). Consequently, the function $\text{RD}^\phi:\mathfrak{F}(T,\mathbb{R}^n)\times T\to \mathbb{R}^n$  is  measurable in its first argument for a fixed $t\in T$. As $T$ is countable and $X$ is a discrete-time stochastic process, it follows that $X(\cdot,\omega)$ is measurable with respect to the product $\sigma$-algebra of Borel $\sigma$-algebras $\mathcal{B}^n$ which is equivalent to the Borel $\sigma$-algebra of $\mathfrak{F}(T,\mathbb{R}^n)$ (see e.g., \citet[Lemma 1.2]{kallenberg1997foundations}). Since function composition preserves measurability, it holds that $\text{RD}^\phi(X(\cdot,\omega),t)$ is measurable in $\omega$ for a fixed $t\in T$. 

%The function $X(\cdot,\omega)$ is measurable in $\omega$ even in the case where $T=\mathbb{R}$ (see e.g., \citet[Prop. 2.6]{capasso2005introduction}).
 
%\textcolor{red}{put in theorem For $\mathcal{RD}^\phi(X(\cdot,\omega),t)$, note that we can write
%\begin{align*}
%\mathcal{RD}^\phi(X(\cdot,\omega),t)&=0.5( 1_{\mathcal{L}^\phi(t)}(X(\cdot,\omega))+1)\text{RD}^{\phi}(X(\cdot,\omega),t)\\
%&\hspace{-0.5cm}+0.5(1_{\mathcal{L}^\phi(t)}(X(\cdot,\omega))-1) \text{RD}^\phi(X(\cdot,\omega),t)
%\end{align*} 
%similarly to \eqref{eq:rho_mu}. While we have shown measurability of $\beta^\phi(X(\cdot,\omega),t)$ in $\omega$ in the proof of Theorem \ref{thm:1},  we can similarly show measurability of $\beta^\phi(X(\cdot,\omega),t)$ in its first argument  with respect to the product $\sigma$ algebra of $\mathfrak{F}(T,\mathbb{R}^n)$. It hence holds that $\mathcal{L}^\phi(t)$ is a measurable set.  Consequently, the functions $1_{\mathcal{L}^\phi(t)}(X(\cdot,\omega))$  and  $\text{RD}^\phi(X(\cdot,\omega),t)$ are measurable in $\omega$ for a fixed $t\in T$. It follows, using similar arguments proceeding \eqref{eq:rho_mu}  in the proof of Theorem \ref{thm:1}, that $\mathcal{RD}^\phi(X(\cdot,\omega),t)$  is measurable in $\omega$ for a fixed $t\in T$.}

\section{Proof of Theorem \ref{thm:2}}
We  prove the statement of Theorem \ref{thm:2} first for the robustness degree $\text{RD}^{\phi}(X,t)$, and finally for the semantics $\beta^\phi(X,t)$, then for the robust semantics  $\rho^\phi(X,t)$. 
 
\subsection{Semantics $\beta^\phi(X,t)$} The proof again follows inductively on the structure of $\phi$. The difference to the proof of Theorem \ref{thm:1} lies in the way  the until operators are handled, which are now assumed to be the non-strict matching versions $\phi' \vec{U}_I \phi''$ and $\phi' \vec{\underline{U}}_I \phi''$. Note also that the time interval $I$ is compact as the formula $\phi$ is assumed to be bounded. The main idea is to show that infimum and supremum operators reduce to minimum and maximum operators that allow us to show measurability. Recall therefore the definition of the future until operator $\beta^{\phi' \vec{U}_I \phi''}(X(\cdot,\omega),t)$ as 
\begin{align*}
	\beta^{\phi' \vec{U}_I \phi''}(X(\cdot,\omega),t)&:=\sup_{t''\in (t\oplus I)\cap T}\Big( \min\big(\beta^{\phi''}(X(\cdot,\omega),t''),\inf_{t'\in[t,t'']\cap T}\beta^{\phi'}(X(\cdot,\omega),t')\big)\Big).
	\end{align*}
	We first show that the infimum operator in $\beta^{\phi' \vec{U}_I \phi''}(X(\cdot,\omega),t)$ reduces to a min operator. In particular, note now that $\underset{t'\in [t,t'']\cap T}{\text{inf}}\beta^{\phi'}(X(\cdot,\omega),t')$ includes the compact time interval $[t,t'']\cap T$ instead of the open interval $(t,t'')\cap T$   due to the interpretation of the until operator as the non-strict matching version. It holds that the minimum of $\underset{t'\in [t,t'']\cap T}{\text{min}}\beta^{\phi'}(X(\cdot,\omega),t')$ exists as
\begin{enumerate}
    \item the minimum is over the compact time interval $[t,t'']\cap T=[t,t'']$ (recall that $T=\mathbb{R}$), and
    \item the range of $\beta^{\phi'}(X(\cdot,\omega),t)$ is restricted to $\mathbb{B}$.
\end{enumerate}
Consequently,  the minimum corresponds to the infimum and it follows that 
\begin{align*}
    \underset{t'\in [t,t'']\cap T}{\text{inf}}\beta^{\phi'}(X(\cdot,\omega),t')=\underset{t'\in [t,t'']\cap T}{\text{min}}\beta^{\phi'}(X(\cdot,\omega),t').
\end{align*}
Now it holds that $\underset{t'\in [t,t'']\cap T}{\text{min}}\beta^{\phi'}(X(\cdot,\omega),t')$ is equivalent to $\beta^{\phi'}(X(\cdot,\omega),t')$ for some $t'\in [t,t'']\cap T$. Since $\beta^{\phi'}(X(\cdot,\omega),t')$ is measurable in $\omega$  by the induction assumption, it follows that the function $\underset{t'\in [t,t'']\cap T}{\text{inf}}\beta^{\phi'}(X(\cdot,\omega),t')$ is measurable in $\omega$ for a fixed $t\in T$. Note next that the supremum operator in $\beta^{\phi' \vec{U}_I \phi''}(X(\cdot,\omega),t)$ reduces to a max operator due to $I$ being compact and following a similar argument as for the infimum operator.  Measurability of $\beta^{\phi' \vec{U}_I \phi''}(X(\cdot,\omega),t)$ in $\omega$  for a fixed $t\in T$ then follow as in the proof of Theorem \ref{thm:1}. The proof for $\beta^{\phi' \vec{\underline{U}}_I \phi''}(X(\cdot,\omega),t)$ follows similarly.

\subsection{Robustness Degree $\text{RD}^{\phi}(X,t)$}
As shown in the proof of Theorem \ref{thm:1}, the function $\text{RD}^\phi:\mathfrak{F}(T,\mathbb{R}^n)\times T\to \mathbb{R}^n$,  is continuous and hence Borel measurable in its first argument for a fixed $t\in T$. By the assumption that $X(\cdot,\omega):\Omega\to\mathfrak{F}(T,\mathbb{R}^n)$ is Borel measureable, the result follows trivially.
 
\subsection{Robust semantics $\rho^\phi(X,t)$} The proof follows mainly from \citep[Theorem 6]{bartocci2015system}. However, to apply this result, we need to show that the robust semantics $\rho^\mu(\zeta,t)$ of predicates $\mu$ are continuous in $\zeta\in\mathbb{R}^n$ where we recall that
\begin{align*}
    \rho^{\mu}(\zeta,t) := \begin{cases} \bar{d}\big(\zeta,\text{cl}(O^{\neg\mu})\big) &\text{if } \zeta\in O^{\mu}\\
	-\bar{d}\big(\zeta,\text{cl}(O^{\mu})\big) &\text{otherwise.}
	\end{cases}
\end{align*}
Note that the functions $\bar{d}\big(\zeta,\text{cl}(O^{\neg\mu})\big)$ and $\bar{d}\big(\zeta,\text{cl}(O^{\mu})\big)$ are continuous in $\zeta$. This follows due to \citet[Chapter 3]{munkres1975prentice}. By definition, we have  $\rho^\mu(\zeta,t)=0$ if $\zeta\in \text{bd}(O^\mu)$ where $\text{bd}(O^\mu)$ denotes the boundary of $O^\mu$. Note also that $\bar{d}\big(\zeta,\text{cl}(O^{\neg\mu})\big)\to 0$ as $\zeta\to \text{bd}(O^\mu)$ as well as $-\bar{d}\big(\zeta,\text{cl}(O^{\mu})\big)\to 0$ as $\zeta\to \text{bd}(O^\mu)$. It  follows that $\rho^{\mu}(\zeta,t)$ is continuous in $\zeta$. The assumption that $X(\cdot,\omega)$ is a cadlag function for each $\omega\in\Omega$ then enables us to apply Theorem 6 in \citet{bartocci2015system}.

\section{Proof of Theorem \ref{thm:3}}

First note that $\rho^\phi(X(\cdot,\omega),t)\le \text{RD}^{\phi}(X(\cdot,\omega),t)$ for each realization $X(\cdot,\omega)$ of the stochastic process $X$ with $\omega\in\Omega$ due to \eqref{underapprox}. Consequently, we have that $ -\text{RD}^{\phi}(X(\cdot,\omega),t)\le -\rho^\phi(X(\cdot,\omega),t)$ for all $\omega\in\Omega$.  If $R$ is now monotone, it directly follows that $R(-\text{RD}^{\phi}(X,t))\le R(-\rho^\phi(X,t))$. 

\section{Proof of Proposition \ref{thm:mmm_}:}

Let us assume that $X^1,\hdots,X^N$ are $N$ independent copies of $X$. Consequently, all $Z^i$ contained within $\mathcal{Z}$ are independent and identically distributed. We first recall the tight version of the Dvoretzky-Kiefer-Wolfowitz inequality as originally presented in~\citet{massart1990tight} which requires that $F_Z$ is continuous.
\begin{lemma}\label{lem:1_}
	Let $\widehat{F}(\alpha,\mathcal{Z})$ be  based on the data $\mathcal{Z}$ consisting of  $Z^1,\hdots,Z^N$ which are $N$ independent copies of $Z$. Let $c>0$ be a desired precision, then it holds that
	\begin{align*}
	P\big(\sup_\alpha|\widehat{F}(\alpha,\mathcal{Z})-{F}_{Z}(\alpha)|>c\big) \le 2\exp\big( -2 Nc^2\big).
	\end{align*}
\end{lemma}

By setting $\delta:=2\exp\big( -2 Nc^2\big)$ in Lemma \ref{lem:1_}, it holds  with a probability of at least $1-\delta$ that
\begin{align*}
	\sup_\alpha|\widehat{F}(\alpha,\mathcal{Z})-{F}_{Z}(\alpha)| \le \sqrt{\frac{\ln(2/\delta)}{2N}}.
	\end{align*}
With a probability of at least $1-\delta$, it now holds that 
\begin{align*}
\Big\{\alpha\in \mathbb{R}|\widehat{F}(\alpha,\mathcal{Z})- \sqrt{\frac{\ln(2/\delta)}{2N}}\ge \beta\Big\}
\subseteq \{\alpha\in \mathbb{R}|{F}_{Z}(\alpha)\ge \beta\}
\end{align*}
as well as
\begin{align*}
\Big\{\alpha\in \mathbb{R}|\widehat{F}(\alpha,\mathcal{Z})+ \sqrt{\frac{\ln(2/\delta)}{2N}}\ge \beta\Big\}
\supseteq \{\alpha\in \mathbb{R}|{F}_{Z}(\alpha)\ge \beta\}.
\end{align*}
Hence, it holds with a probability of at least $1-\delta$ that
\begin{align*}
\inf\Big\{\alpha\in \mathbb{R}|\widehat{F}(\alpha,\mathcal{Z})
 - \sqrt{\frac{\ln(2/\delta)}{2N}}\ge \beta\Big\} 
\ge \inf\{\alpha\in \mathbb{R}|{F}_{Z}(\alpha)\ge \beta\}
\end{align*}
as well as 
\begin{align*}
\inf\Big\{\alpha\in \mathbb{R}|\widehat{F}(\alpha,\mathcal{Z})
 + \sqrt{\frac{\ln(2/\delta)}{2N}}\ge \beta\Big\}
 \le \inf\{\alpha\in \mathbb{R}|{F}_{Z}(\alpha)\ge \beta\}.
\end{align*}
By the definition of $\underline{VaR}_\beta(\mathcal{Z},\delta)$ and $\overline{VaR}_\beta(\mathcal{Z},\delta)$, it holds with a probability of at least $1-\delta$ that
\begin{align*}
    \underline{VaR}_\beta(\mathcal{Z},\delta)\le VaR_\beta(Z)\le \overline{VaR}_\beta(\mathcal{Z},\delta).
\end{align*}  

\section{Proof of Proposition \ref{thm:mmmmm}}		
Let us again assume that $X^1,\hdots,X^N$ are $N$ independent copies of $X$. Consequently, all $Z^i$ contained within $\mathcal{Z}$ are independent and identically distributed. Note first that $\widehat{E}(\mathcal{Z})$ is a random variable with the expected value according to
\begin{align*}
E(\widehat{E}(\mathcal{Z}))&=\frac{1}{N}\sum_{i=1}^N E(Z_i)=\frac{1}{N}\sum_{i=1}^N E(Z)=E(Z).
\end{align*}
For $c> 0$, we can now apply Hoeffding's inequality and obtain the concentration inequality
\begin{align*}
P\big(|\widehat{E}(\mathcal{Z})-E(Z)|\ge c\big)&\le 2\exp\Big(-\frac{2Nc^2}{(b-a)^2}\Big).
\end{align*}
By setting $\delta:=2\exp\Big(-\frac{2Nc^2}{(b-a)^2}\Big)$, it holds  with a probability of at least $1-\delta$ that
\begin{align*}
	|\widehat{E}(\mathcal{Z})-E(Z)| \le \sqrt{\frac{\ln(2/\delta)(b-a)^2}{2N}}.
\end{align*}
From this inequality, the result follows trivially.

\section{Discretization of $c$ and $d$ in Example \ref{ex1}}
\label{app:discretization}
To discretize the distributions of $c$ and $d$ in \eqref{distr_c} and \eqref{distr_d}, respectively, let $M:=32$ be the number of desired discretization steps and $\gamma:=0.55$ be a discretization bound. We uniformly discretize the interval $[-\gamma,\gamma]$ into $M$ values $(s_1,\hdots,s_M)$ where $s_m< s_{m+1}$. We additionally add $s_0:=0$ and define $S:=(s_0,s_1,\hdots,s_M)$. We now assign a PMF $f_S(s_m)$ to each element $s_m\in S$ as
\begin{align*}
    f_S(s_m):=\begin{cases}
        F_\mathcal{N}(s_m)    & \text{if } s_m=s_1\\
        F_\mathcal{N}(s_m)- F_\mathcal{N}(s_{m-1}) & \text{if } s_1<s_m<0\\
        2(F_\mathcal{N}(s_m)- F_\mathcal{N}(s_{m-1})) & \text{if } s_m=0\\
        F_\mathcal{N}(s_{m+1})- F_\mathcal{N}(s_m) & \text{if } 0<s_m<s_M\\
        1- F_\mathcal{N}(s_m) & \text{if } s_m=s_M\\
        \end{cases}
\end{align*}
where $F_\mathcal{N}(s)$ is the CDF of $\mathcal{N}(0,0.2)$ (according to \eqref{distr_c} and \eqref{distr_d}). We now assume, instead of \eqref{distr_c} and \eqref{distr_d}, that $c$ and $d$ take values in the sets
 \begin{align*}
     \mathcal{C}:=2\oplus S \times 3 \oplus S\\
     \mathcal{D}:=6\oplus S \times 4 \oplus S
 \end{align*}
 where $2$, $3$, $6$, and $4$ are the mean values of $c$ and $d$ in \eqref{distr_c} and \eqref{distr_d}, respectively. Finally, we assume that the distributions of $c=\begin{bmatrix}c_1 & c_2 \end{bmatrix}^T$ and $d=\begin{bmatrix}d_1 & d_2 \end{bmatrix}^T$ are according to the PMFs $f_c(c):=f_S(c_1)f_S(c_2)$ and $f_d(d):=f_S(d_1)f_S(d_2)$, respectively.